\documentclass[preprint,nofootinbib,3p,showpacs]{elsarticle}
\usepackage{graphicx}
\usepackage{amsmath}
\usepackage{amssymb}
\usepackage{overpic,subfigure}
\usepackage{multirow}
\usepackage[symbol]{footmisc}
\usepackage{dcolumn}
\usepackage{bm}
\usepackage{rotating}
\usepackage{hyperref}
\hypersetup{colorlinks=true}
\usepackage{indentfirst}
\usepackage{lineno}
\usepackage{epstopdf}
\usepackage{units}
\usepackage{amsthm}
\usepackage{url}
\usepackage{amsfonts}
\usepackage{textcomp}
\usepackage{multicol}
\usepackage{verbatim}
\usepackage{rotating}
\usepackage{float}
\usepackage{color}
\usepackage{pstricks}
\usepackage{pst-node}
\usepackage{times}
\usepackage{ulem}
\usepackage{booktabs}
\usepackage{multicol}
\usepackage[mathscr]{eucal}

\newcommand{\BESIIIorcid}[1]{\href{https://orcid.org/#1}{\hspace*{0.1em}\raisebox{-0.45ex}{\includegraphics[width=1em]{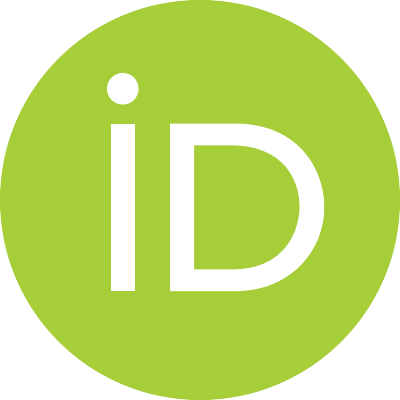}}}}


\def \ee   {e^+e^-}

\newcommand{\bfg}{\begin{figure}}
\newcommand{\efg}{\end{figure}}
\newcommand{\bitm}{\begin{itemize}}
\newcommand{\eitm}{\end{itemize}}
\newcommand{\bnum}{\begin{enumerate}}
\newcommand{\enum}{\end{enumerate}}
\newcommand{\btbl}{\begin{table}}
\newcommand{\etbl}{\end{table}}
\newcommand{\btbu}{\begin{tabular}}
\newcommand{\etbu}{\end{tabular}}

\newcommand{\beq}{\begin{equation}}
\newcommand{\edq}{\end{equation}}

\begin{document}

\begin{frontmatter}

\title{\boldmath Search for the charged lepton flavor violating decay $\psi(3686)\to e^{\pm}\mu^{\mp}$}

\author{BESIII collaboration\fnref{authors} \corref{email}}
\fntext[authors]{Authors are listed at the end of this paper.}
\cortext[email]{Corresponding email: besiii-publications@ihep.ac.cn}

\date{\today}

\begin{abstract}

By analyzing $(2367.0\pm11.1)\times10^6$ $\psi(3686)$ events collected in $e^+e^-$ collisions at $\sqrt{s}=3.686~\rm GeV$ with the BESIII detector at the BEPCII collider, we report the first search for the charged lepton flavor violating decay $\psi(3686)\to e^{\pm}\mu^{\mp}$. No signal is found. An upper limit on the branching fraction $\mathcal{B}(\psi(3686)\to e^{\pm}\mu^{\mp})$ is determined to be $1.4\times10^{-8}$ at the 90\% confidence level.
  
\end{abstract}

\end{frontmatter}


%
%
\section{Introduction}
\label{sec:introduction}
In the Standard Model~(SM), quarks change flavor through electroweak transitions, with rates codified by the Cabibbo-Kobayashi-Maskawa matrix.  Similarly, neutrinos oscillate between electron, muon, and tau lepton flavors according to rates given by the Pontecorvo-Maki-Nakagawa-Sakata matrix, which is beyond SM~\cite{ref::neutrino_1}. However, the lepton flavor violation has not been discovered in the sector of charged leptons. The discovery of non-zero neutrino masses and neutrino oscillations guarantees that charged lepton flavor violation~(CLFV) can occur in SM loops, but with very small branching fractions~(BFs), which are experimentally undetectable. Therefore, any detection of CLFV would be an unambiguous signal of new physics beyond the SM~\cite{ref::clfvreview_1}.

Over the past several decades, searches for CLFV processes have been performed in the lepton, pseudo-scalar meson, vector meson, and boson sectors~\cite{ref::CLFVreview}. For example, in the lepton sector, searches for CLFV have been carried out in muon and tau decays, such as $\mu^{+}\to e^+\gamma$~\cite{ref::mu2egamma}, $\mu^{+}\to e^{+}e^{-}e^{+}$~\cite{ref::mu2eee},  $\mu^{-}N\to e^{-}N$~\cite{ref::muN2eN_1,ref::muN2eN_2}, $\mu^{-}N\to e^{-}N'$~\cite{ref::muN2eNprime}, $\tau^{\pm}\to l^{\pm}\gamma$~\cite{ref::tau2lgamma}, $\tau\to lll$~\cite{ref::tau2lll_1,ref::tau2lll_2}, and $\tau\to hl$~\cite{ref::tau2hl_1,ref::tau2hl_2} (where $l$ denotes a lepton and $h$ denotes a hadron, such as the $\pi^0$, $\eta^{(\prime)}$, $\phi$, $\omega$, $K^{*0}$, or $\rho^0$).
However, no evidence of CLFV has yet been found.
In the quarkonium sector, several CLFV decays of the $J/\psi$ and $\Upsilon(nS)$ have been investigated by the BESIII, BaBar, and Belle experiments, and no evidence for a signal has been found. Upper limits~(UL) on the BFs of the decays $J/\psi\to e\mu$, $J/\psi\to e\tau$, $\Upsilon(1S)\to e\mu$, $\Upsilon(1S)\to \mu\tau$, $\Upsilon(1S)\to e\tau$, and $\Upsilon(3S)\to e\mu$ are currently $4.5\times10^{-9}$~\cite{ref::jpsi2emu}, $7.5\times10^{-8}$~\cite{ref::jpsi2etau}, $3.6\times10^{-7}$, $2.6\times10^{-6}$, $2.4\times10^{-6}$~\cite{ref::upsilon2ll_1}, and $3.6\times10^{-7}$~\cite{ref::upsilon2ll_2} at $90\%$ confidence level~(C.L.), respectively.

Theoretically, CLFV can be described by various new physics models, such as the two-Higgs doublet model with extra Yukawa couplings~\cite{ref::yukawa}, a model that constitutes a simple example of tree-level off-diagonal Majorana couplings not suppressed by neutrino masses~\cite{ref::majorana}, and a series of models based on super-symmetry (SUSY)~\cite{ref::susy_gut, ref::susy_leptons, ref::susy_neutrino, ref::susy_zprime}.  For the CLFV decays of charmonium, models involving SUSY-based grand unified theories~\cite{ref::clfvpsip_gut}, SUSY~\cite{ref::clfvpsip_susy}, and TC2~\cite{ref::clfvpsip_tc2} with various new bosons (scalars, vectors), predict their decay rates at the order of $10^{-8}$ to $10^{-16}$~\cite{ref::theory_clfvpsip_1,ref::theory_clfvpsip_2,ref::theory_clfvpsip_3,ref::theory_clfvpsip_4,ref::theory_clfvpsip_5,ref::theory_clfvpsip_6}.

%
%
The BESIII analysis reported in this Letter is the first search for CLFV in the decay $\psi(3686)\to e^{\pm}\mu^{\mp}$. In this Letter, $(2367.0\pm 11.1)\times 10^{6}$ $\psi(3686)$ events collected in 2009 and 2021 at $\sqrt{s}=3.686~\rm GeV$~\cite{ref:numpsip} are analyzed. 
The $e^{+}e^{-}$ collision events collected at $\sqrt{s}=3.773$, $3.65$, and $3.682~\rm GeV$~\cite{ref:data-3650-3773,ref:data-3773,Liao:2025lth}, corresponding to the integrated luminosities of $7926.8$, $454.33$, and $404~\rm pb^{-1}$, respectively, are used to estimate the continuum background. 
%
%

\section{BESIII Detector and data samples}

The BESIII detector~\cite{Ablikim:2009aa} records symmetric $e^+e^-$ collisions provided by the BEPCII storage ring~\cite{Yu:IPAC2016-TUYA01} in the center-of-mass~(c.m.) energy range from 1.84 to 4.95~GeV, with a peak luminosity of $1.1 \times 10^{33}\;\text{cm}^{-2}\text{s}^{-1}$ achieved at $\sqrt{s} = 3.773\;\text{GeV}$. BESIII has collected large data samples in this energy region.  The data is monitored in real time using the Bhabha process and the Data Quality Monitoring system~\cite{EcmsMea}. The cylindrical core of the BESIII detector covers 93\% of the full solid angle and consists of a helium-based multilayer drift chamber~(MDC), a time-of-flight system~(TOF), and a CsI(Tl) electromagnetic calorimeter~(EMC), which are all enclosed in a superconducting solenoidal magnet providing a 1.0~T magnetic field.
The solenoid is supported by an
octagonal flux-return yoke with resistive plate muon counters~(MUC)
identification modules interleaved with steel. 
The charged-particle momentum resolution at $1~{\rm GeV}/c$ is
$0.5\%$, and the 
${\rm d}E/{\rm d}x$
resolution is $6\%$ for electrons
from Bhabha scattering. The EMC measures photon energies with a
resolution of $2.5\%$ ($5\%$) at $1$~GeV in the barrel (end cap)
region. The time resolution in the plastic scintillator TOF barrel region is 68~ps, while that in the end cap region was upgraded in 2015 from 110~ps to 60~ps using multi-gap resistive plate chamber technology~\cite{ref:tofupgrade}. About 95\% of the dataset used in this analysis benefits from this upgrade. 


%
%
Monte Carlo~(MC) simulated data samples produced with a {\sc
geant4}-based~\cite{geant4} software package, which
includes the geometric description~\cite{Huang:2022wuo, Song:2025pnt} of the BESIII detector and the
detector response, are used to determine detection efficiencies
and to estimate backgrounds. The simulation models the beam
energy spread and initial state radiation (ISR) in the $e^+e^-$
annihilations with the generator {\sc
kkmc}~\cite{ref:kkmc1,ref:kkmc2}.
The inclusive MC sample includes the production of the
$\psi(3686)$ resonance, the ISR production of the $J/\psi$, and
the continuum processes incorporated in {\sc
kkmc}.
All particle decays are modelled with {\sc
evtgen}~\cite{ref:evtgen1,ref:evtgen2} using BFs 
either taken from the
Particle Data Group~\cite{ref:pdg}, when available,
or otherwise estimated with {\sc lundcharm}~\cite{ref:lundcharm1,ref:lundcharm2}.
Final state radiation
from charged final state particles is incorporated using the {\sc
photos} package~\cite{photos2}.
Samples of $1\times10^5$ events are generated separately for the experimental conditions of 2009 and 2021 to study the selection criteria. Approximately $2.406\times10^9$ $\psi(3686)$ events from the inclusive MC sample are used to estimate the backgrounds from $\psi(3686)$ decays. To improve the statistics of the MC sample for background studies, an equivalent of $2.406\times10^{10}$ $\psi(3686)$ exclusive events are generated, including decay channels such as $\psi(3686)\to e^{+}e^{-},~\mu^{+}\mu^{-},~\tau^+\tau^-,~K^+K^-$, and $p\bar{p}$.
To avoid involuntary bias, a semi-blind analysis method is employed. The full dataset is analyzed after the analysis procedure is finalized by using the MC simulation sample and $10\%$ of the full data sample.
\section{Event Selection}

Signal events are reconstructed with an electron and a muon that are back-to-back in the rest frame of the $\psi(3686)$. 
All charged tracks are required to be within a polar angle range of $\vert\!\cos\theta\vert<0.93$, where $\theta$ is the polar angle of charged tracks. The good charged tracks must originate from the interaction region, defined by $|V_{xy}|<1~\rm cm$ and $|V_z|<10~\rm cm$, where $|V_{xy}|$ and $|V_{z}|$ are the distances of closest approach of the reconstructed track to the interaction point in the $xy$ plane and the $z$ direction (along the beam), respectively. For the signal channel, the number of good charged tracks is required to be two, and the net charge of these two tracks is required to be zero. To reject cosmic rays, the TOF difference of the two good charged tracks must be less than $1.0~\rm ns$.

\begin{figure*}[htbp]
\centering
\includegraphics[width=0.45\textwidth]{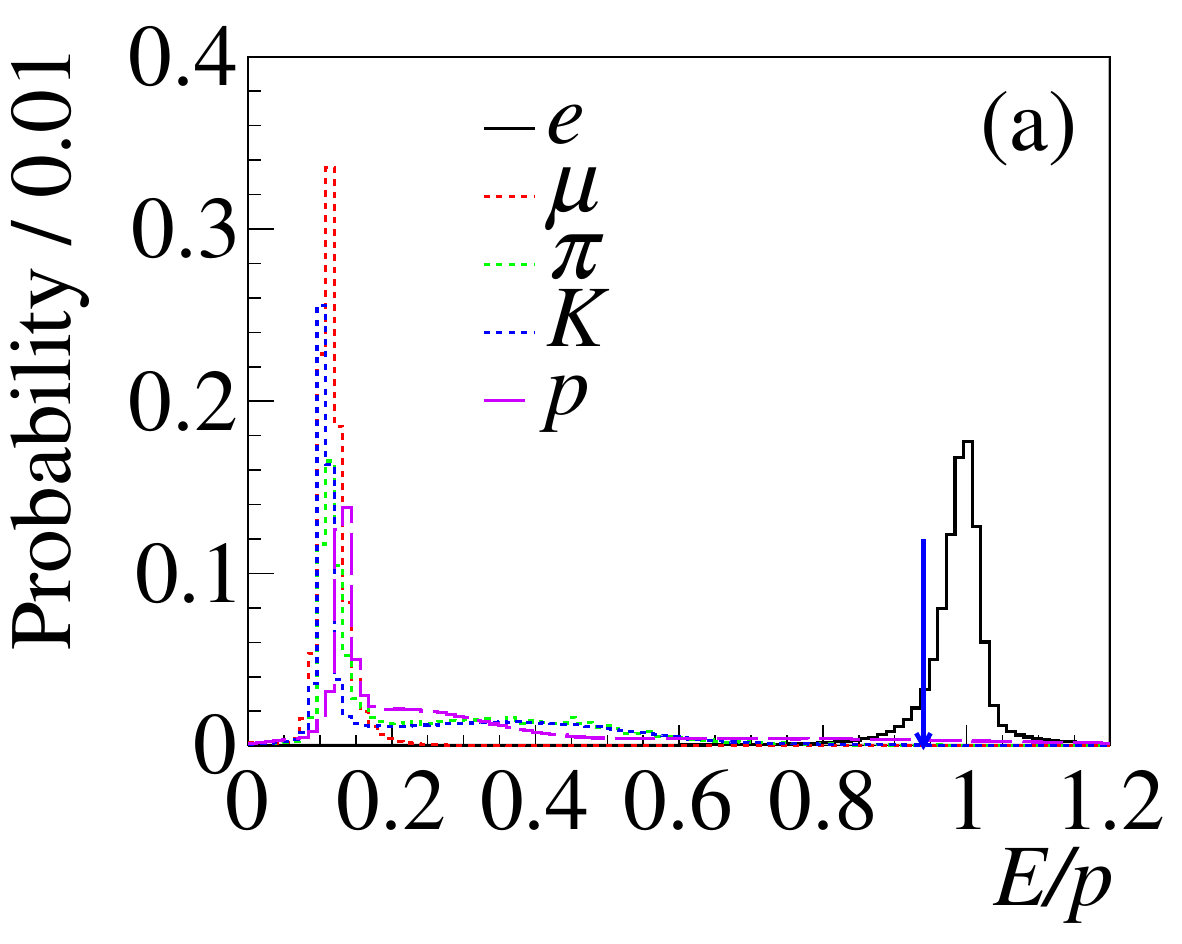}
\hspace{0.1cm}
\includegraphics[width=0.45\textwidth]{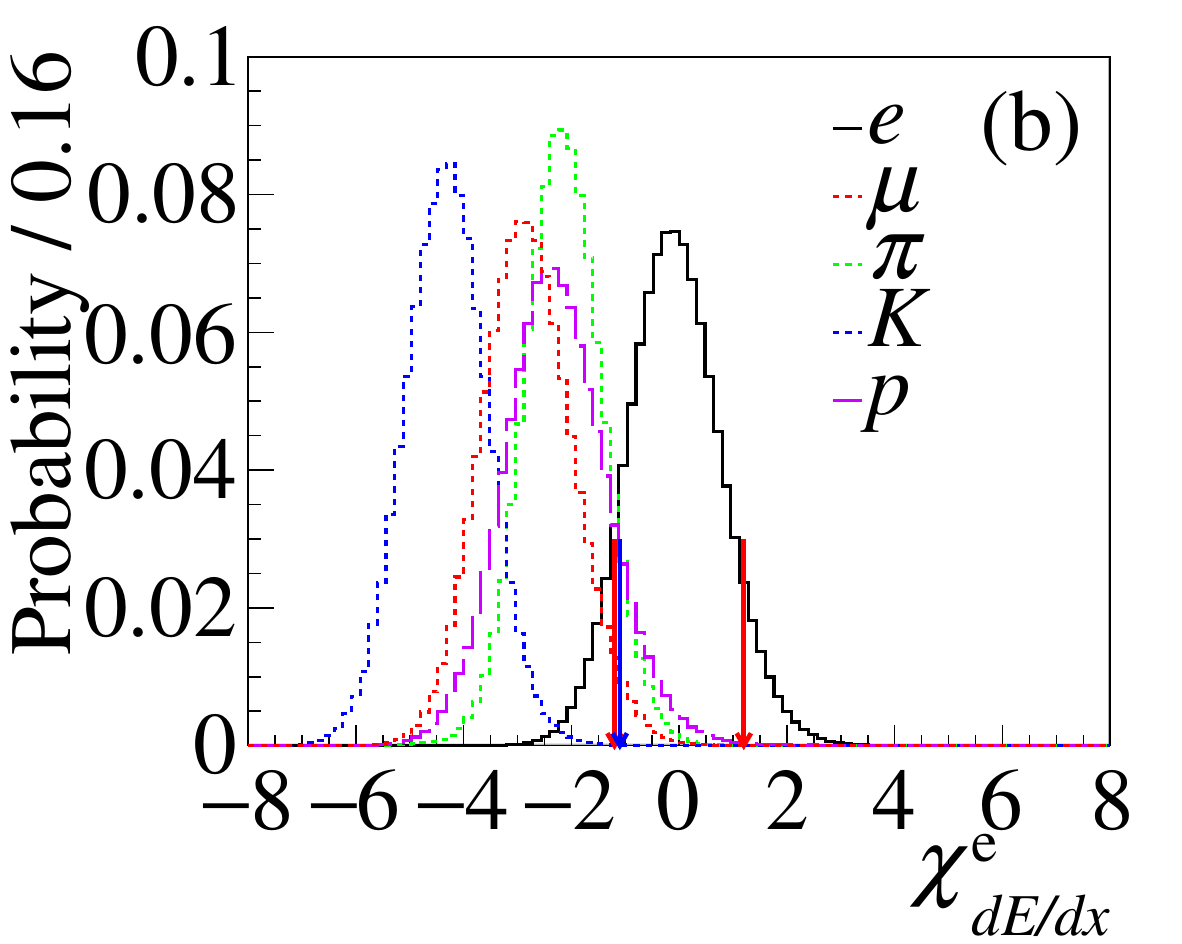}
\caption{
    Distributions of $E/p$ (a) and $\chi^{e}_{dE/dx}$  (b) for $e$, $\mu$, $\pi$, $K$, and $p$ samples from MC simulations of $\psi(3686)\to e^+e^-$, $\mu^+\mu^-$, $\pi^+\pi^-$, $K^+K^-$, and $p\bar{p}$, respectively. In (a), the blue arrow indicates $E/p>0.94c$ for electron selection. In (b), the red arrows indicate $-1.2<\chi^{e}_{dE/dx}<1.2$ for electron selection, and the blue arrow indicates $\chi^{e}_{dE/dx}<-1.1$ for muon selection.
  }
\label{fig::dis_ededxe_epratio}
\end{figure*}

\begin{figure}[htbp]
\centering
\includegraphics[width=0.45\textwidth]{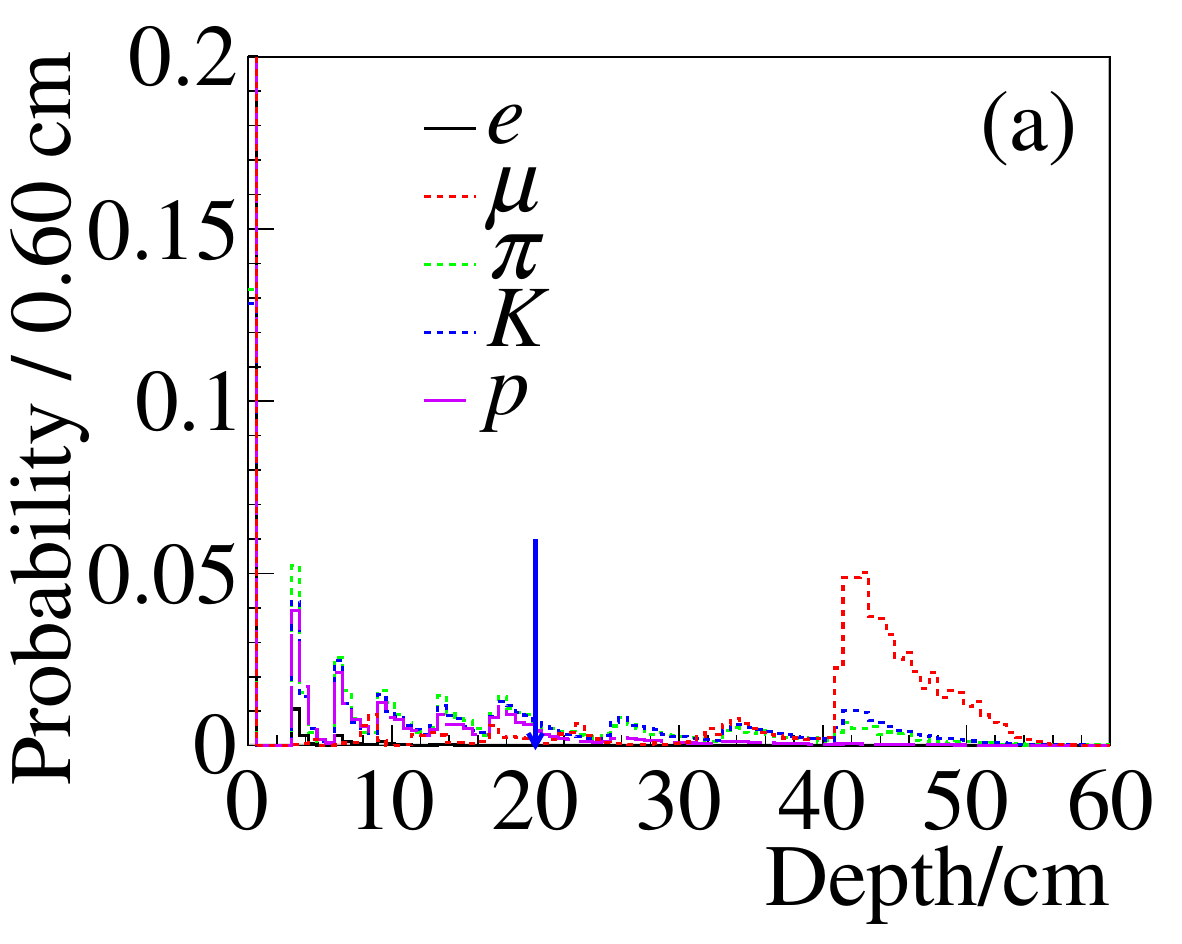}
\hspace{0.1cm}
\includegraphics[width=0.45\textwidth]{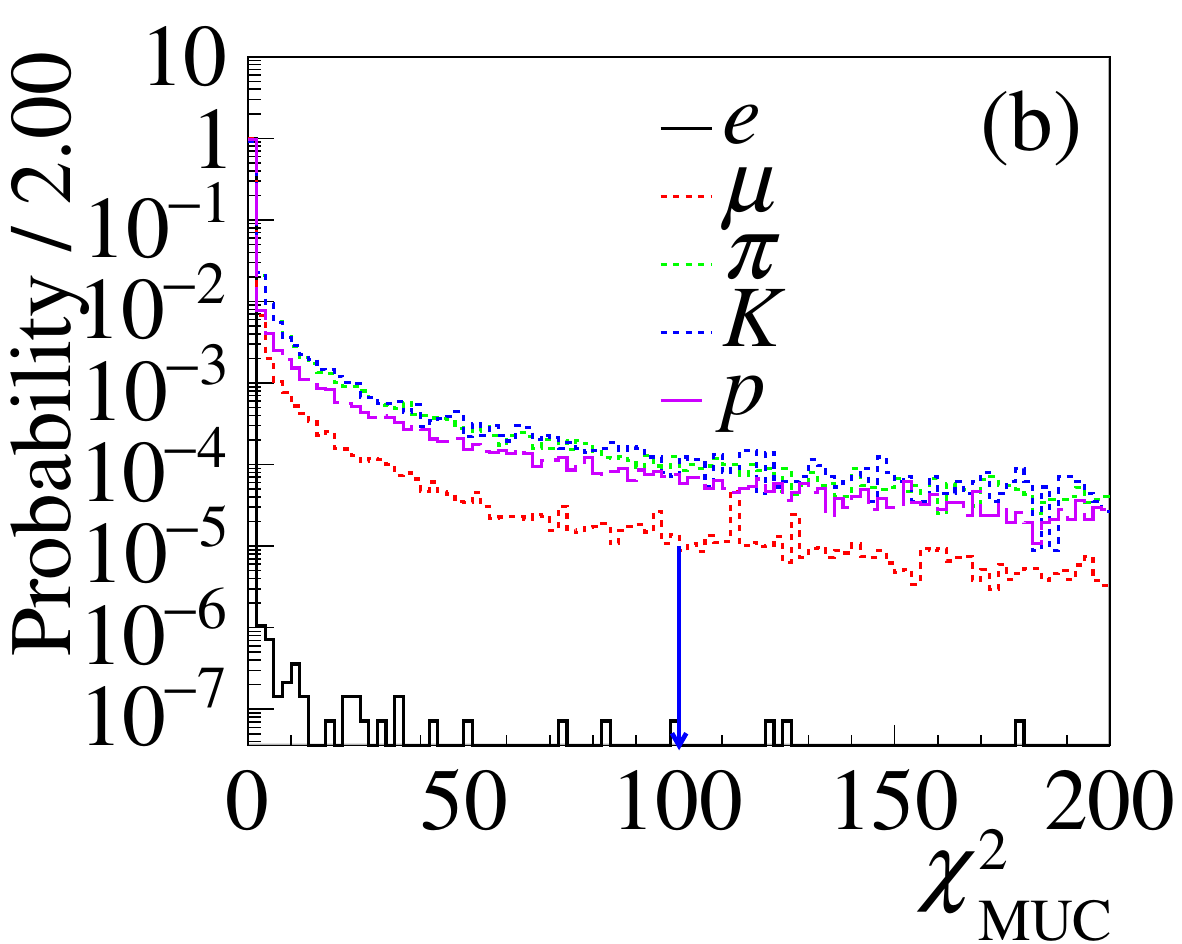}
\caption{
    Distributions of the MUC depth (a) and $\chi^2_{\rm MUC}$ (b) for different particle samples from MC simulations. The blue arrows indicate the muon selection criteria: MUC depth $> 20~\rm cm$ in (a) and $\chi^2_{\rm MUC}<100$ in (b).
  } 
\label{fig::dis_depth_chi2muc}
\end{figure}

Strict particle identification~(PID) for the electron and muon is based on information provided by the MDC, EMC and MUC sub-detectors.
For electron PID, the energy-to-momentum ratio $E/p$ is required to be greater than $0.94c$, as shown in Fig. \ref{fig::dis_ededxe_epratio}(a), where $E$ is the deposited energy in the EMC and $p$ is the magnitude of the momentum measured by the MDC. In addition, $\chi^{e}_{dE/dx}$ is required to satisfy $|\chi^{e}_{dE/dx}|<1.2$, as shown in Fig. \ref{fig::dis_ededxe_epratio}(b), where $\chi^{e}_{dE/dx}$ is the difference between the measured and expected specific ionization energy loss $dE/dx$ under the electron hypothesis, normalized by the $dE/dx$ resolution. Furthermore, since electrons cannot pass through the EMC and reach the MUC, the maximum number of hits in any layer of the MUC for the electron is required to be zero.
For muon PID, an energy window of $[0.1,~0.3]~\rm GeV$ is applied, where the energy is the deposited energy measured by the EMC. In order to lower the probability to misidentify pions or kaons as muons, the penetration depth of the muon in the MUC is required to be greater than $20~\rm cm$, as shown in Fig. \ref{fig::dis_depth_chi2muc}(a). To obtain a well-reconstructed MUC track, the number of layers penetrated by the muon in the MUC must be greater than three, and the fitted $\chi^2_{\rm MUC}$ for the muon in the MUC must be less than 100, as shown in Fig. \ref{fig::dis_depth_chi2muc}(b). Additionally, to suppress misidentification due to electrons, the value of $\chi^{e}_{dE/dx}$ for the muon is required to be less than $-1.1$, as shown in Fig. \ref{fig::dis_ededxe_epratio}(b).

The energy leakage at $|\cos\theta| = 0$ causes the rising of the rate of the misidentification from electron to muon. To further suppress the background from Bhabha scattering events, in which one of the electrons is misidentified as a muon, additional selections are applied to the muon candidates. Muon candidates with an MDC track polar angle satisfying $0.82<\vert\!\cos\theta\vert<0.86$ are vetoed because of the gap of barrel and endcap of EMC. The track direction difference of the muon candidate in the MDC and MUC is used to constrain background. The polar angle difference between the track in the MUC and MDC is required to be less than $20^{\circ}$, and the azimuthal angle difference between the track in the MUC and MDC is required to be less than $30^{\circ}$. A detailed discussion can be found in Ref.~\cite{ref:visualbes3}.

Since the final state electron and muon are back-to-back, the opening angles in the polar and azimuthal directions are required to be less than $1.4^{\circ}$ and $1.7^{\circ}$, respectively. The opening angle in the polar direction is defined as $|\Delta\theta|=|180^{\circ}-(\theta_1+\theta_2)|$, where $\theta_1$ and $\theta_2$ are the corresponding polar angles of the electron and muon, respectively. The opening angle in the azimuthal direction is defined as $|\Delta\phi|=|180^{\circ}-|\phi_1-\phi_2||$, where $\phi_1$ and $\phi_2$ are the azimuthal angles of the electron and muon, respectively.

To suppress background from the processes $e^+e^-\to \gamma e^+e^-$ and $e^+e^-\to \gamma\mu^+\mu^-$,
the number of detected photons is required to be zero  
The photon candidates used for this veto are reconstructed using showers in the EMC detector. The showers are required to have a deposited energy greater than $25~\rm MeV$ in the barrel region ($\vert\!\cos\theta\vert<0.80$) and $50~\rm MeV$ in the end-cap region ($0.86<\vert\!\cos\theta\vert<0.92$). The difference between the EMC time and the event start time for each photon must be within [0, 700]~ns to suppress electronic noise and showers unrelated to the event. To exclude showers originating from the final state charged tracks, the opening angle between the position of each shower in the EMC and the closest extrapolated charged track must be greater than 10 degrees.

The normalized momentum sum $|\sum_{i}\vec{p_{i}}|/\sqrt{s}$ and the normalized energy sum $\sum_{i} E_{i}/\sqrt{s}$ are used to determine the number of signal events in the full dataset, where $\vec{p}_{i}$ is the momentum and $E_{i}$ is the reconstructed energy of final state particle $i$, and $\sqrt{s}$ is the c.m. energy. The signal region is defined by $|\sum_{i}\vec{p_{i}}|/\sqrt{s}\leq 0.03$ and $0.97\leq\sum_{i} E_{i}/\sqrt{s}\leq1.04$.
The selection criteria discussed above are optimized by maximizing the figure-of-merit (FOM) suggested by Punzi~\cite{ref:punzi}, given by

\begin{equation}
    {\rm FOM}=\frac{\epsilon^{\rm MC}_{\rm signal}}{a/2+\sqrt{N_{\rm exp}}}.
\end{equation}
Here, $\epsilon^{\rm MC}_{\rm signal}$ is the detection efficiency of the signal channel estimated by the signal MC sample, $a$ is the significance value, which is set to 3, and $N_{\rm exp}$ is the number of expected background events in the full dataset.

\begin{figure*}[htbp]
\centering
\includegraphics[width=0.45\textwidth]{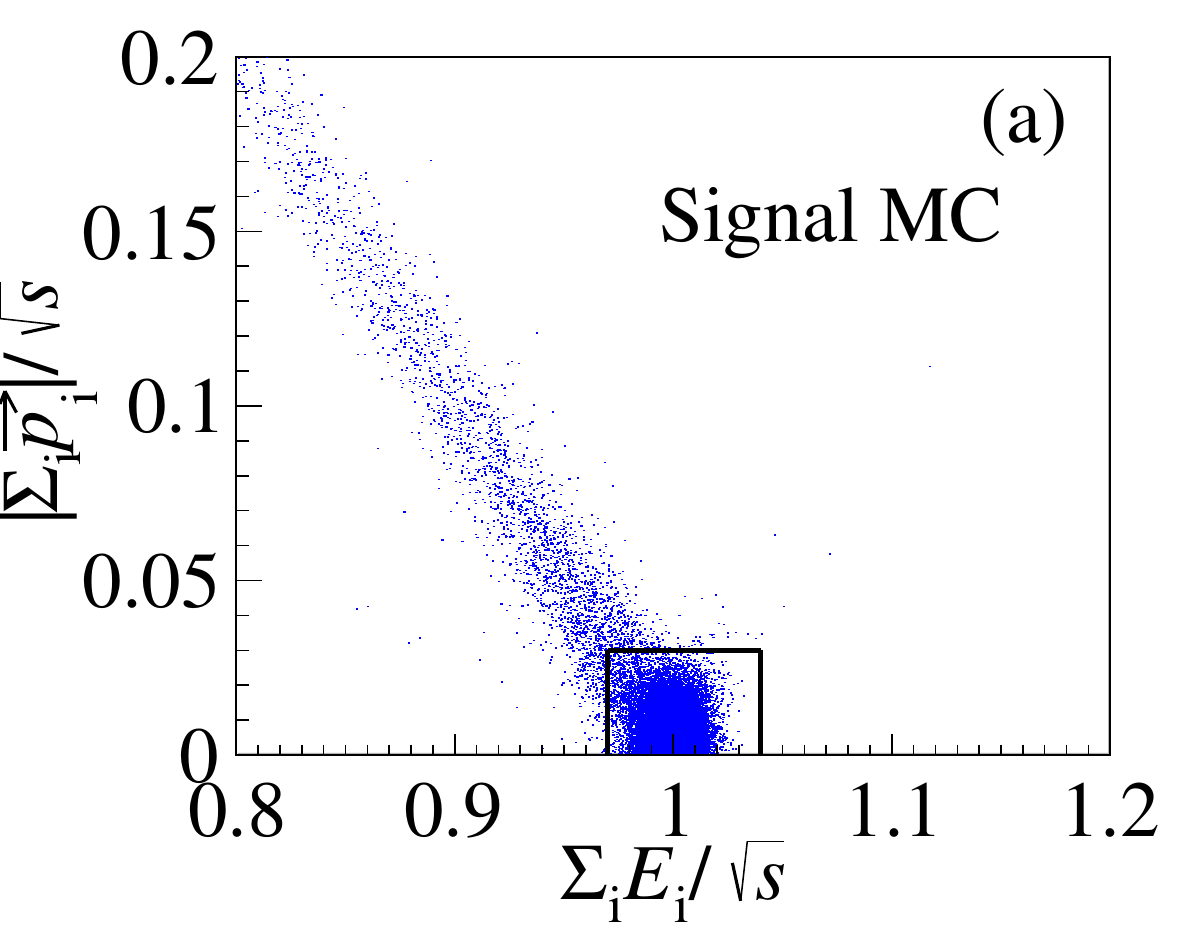}
\hspace{0.1cm}
\includegraphics[width=0.45\textwidth]{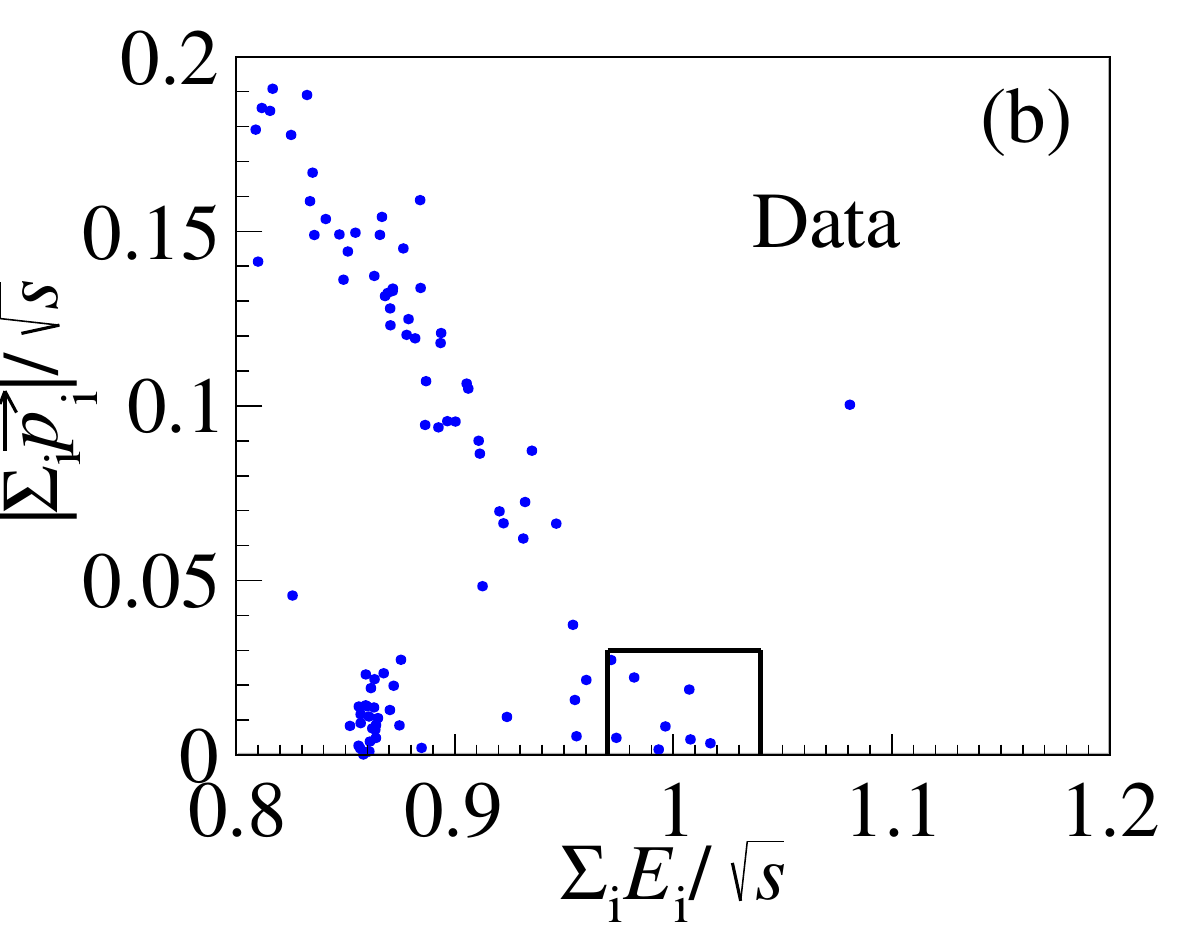}
\caption{
  The distribution of $|\sum_{i}\vec{p_{i}}|/\sqrt{s}$ versus $\sum_{i} E_{i}/\sqrt{s}$ for the signal MC sample (a) and the full $\psi(3686)$ data (b). The indicated signal region is defined as $|\sum_{i}\vec{p_{i}}|/\sqrt{s}\leq 0.03$ and $0.97\leq\sum_{i} E_{i}/\sqrt{s}\leq1.04$.
  }
\label{fig::signal_box}
\end{figure*}

%
%
The detection efficiency for the signal process is estimated using the signal MC sample, which is determined to be $(24.18\pm0.10)\%$. The distribution of the signal MC sample inside and outside the signal region are shown in Fig \ref{fig::signal_box}(a). After applying all the selections to the data from $2367.0\times10^6$ $\psi(3686)$ events, eight $\psi(3686)\to e^{\pm}\mu^{\mp}$ candidates are observed in the signal region, as shown in Fig. \ref{fig::signal_box}(b).

%
%
\section{Background study}

There are two types of background. One comes from $\psi(3686)$ decaying into two charged particles, including $\psi(3686)\to e^+e^-$, $\mu^+\mu^-$, $\tau^+\tau^-$, $\pi^+\pi^-$, $K^+K^-$, and $p\bar{p}$. The other type is the continuum background from scattering processes, including  $e^+e^-\to e^+e^-(\gamma)$ and $e^+e^-\to \mu^+\mu^-(\gamma)$. 

The number of background events from $\psi(3686)$ decay is estimated by analyzing the $\psi(3686)$ inclusive and exclusive MC samples. The equivalent statistics of the inclusive MC sample is about the same as the data and the exclusive MC sample is ten times larger. The scaled number of background events from $\psi(3686)$ decay in the signal region is estimated to be $1.6\pm0.4$.


The number of background events from the continuum process is estimated using $e^+e^-$ collision data samples at surrounding energy points, such as $\sqrt{s}=3.65,~3.682$ and $3.773~\rm GeV$. The c.m. energy and cross section of the continuum process have a inverse square relationship. Therefore, the scaled number of continuum background events at the $\psi(3686)$ c.m. energy can be obtained from the different data samples by
\begin{equation}
    N^{s,k}_{\rm bkg2}=N^{k}_{\rm cont}\times f^{k}_{2},~f^{k}_{2}=\frac{\mathscr{L}_{\psi(3686)}}{\mathscr{L}_k}\times \frac{s_k}{s_{\psi(3686)}},
\end{equation}
where $N^{k}_{\rm cont}$ is the number of background events that survive in the signal region of the data sample with c.m. energy corresponding to index $k$, while $\mathscr{L}_k$ and $\mathscr{L}_{\psi(3686)}$ are the integrated luminosities of the data sample at c.m. energy $k$ and at the $\psi(3686)$ peak energy, respectively. $\sqrt{s_k}$ and $\sqrt{s_{\psi(3686)}}$ are the c.m. energies of different data samples. By weighting with errors, the scaled number of background events from the different data samples is combined to be $N^{s}_{\rm bkg2}=\sum_{k}(N^{s,k}_{bkg2}/\sigma^2_k)/\sum_k(1/\sigma^2_k)=4.6\pm1.4$, where $\sigma_k$ is the error of $N^{s,k}_{bkg2}$. The expected sum of the background events from $\psi(3686)$ decay and the scaled continuum production is estimated to be $6.2\pm1.4$.


%
%
\section{Systematic uncertainties}

The systematic uncertainty primarily arises from the tracking and PID of electrons and muons, the TOF difference requirement, the photon veto, the $|\Delta\theta|$ and $|\Delta\phi|$ requirement, the Bhabha event veto, the signal region requirement, the signal MC generation model, the statistics of the signal MC sample, and the number of $\psi(3686)$ events in data. The uncertainties from all sources are listed in Table~\ref{tab:syst}. Control samples of $\psi(3686)(e^+e^-)\to e^+e^-$ and $\psi(3686)(e^+e^-)\to \mu^+\mu^-$ decays are used to estimate the systematic uncertainties from different selection requirements.
The differences between MC and data for these two control samples give the uncertainties from tracking and PID selections. We combine these uncertainties by comparing weighted and unweighted signal efficiency. The weighted signal efficiency is defined as $\epsilon_{sig}^{w} = \sum_{i=1}^{n}\left(\prod_{r}\left(\epsilon_r^{\rm~Data}(P_{(t)},\cos\theta)/ \epsilon_r^{\rm~MC}(P_{(t)},\cos\theta)\right)/n\right)$, where $\epsilon_{sig}^{w}$ is the weighted efficiency, $\epsilon_r^{\rm~Data}(P_{(t)},\cos\theta)$ and $\epsilon_r^{\rm~MC}(P_{(t)},\cos\theta)$ are the efficiency of control samples at different momentum and polar angle region, respectively, $r$ denotes different selection criteria, $n$ is the number of events in control samples. The systematic uncertainty of tracking and PID for the electron and muon is $0.5\%$.
The uncertainties from other requirements are estimated by comparing the efficiencies in MC and data for the two control samples, where the larger efficiency difference is taken as the systematic uncertainty for each selection criterion.
The uncertainties from the TOF difference requirement, photon veto, $\Delta\theta$ and $\Delta\phi$ requirements, Bhabha events veto selection, and signal region requirement are determined to be $0.2\%$, $1.1\%$, $3.8\%$, $3.3\%$, and $1.7\%$, respectively.
The uncertainty from the signal MC model is estimated by comparing signal MC samples generated with two different models, which are the VLL model and a uniform phase-space model~\cite{ref:evtgen1}, respectively. The uncertainty of the signal MC model is $10.0\%$.
The uncertainty from signal MC statistics is determined to be $0.4\%$ according to the number of events in the signal MC sample. 
The uncertainty in the number of $\psi(3686)$ events from different years is quoted as $0.5\%$~\cite{ref:numpsip}.
Assuming no correlation between these sources, the total systematic uncertainty is $11.4\%$.

\begin{table*}[htbp] 
\setlength{\abovecaptionskip}{0.0cm}
\setlength{\belowcaptionskip}{-1.6cm}
\centering
\caption{Summary of systematic uncertainties from all considered sources.}
\begin{center}
      \footnotesize
      \newcommand{\tabincell}[2]{\begin{tabular}{@{}#1@{}}#2\end{tabular}}
    \begin{tabular}{cc}
        \hline \hline
        
        Source & Uncertainty(\%)  \\
        \hline
        
        Tracking and PID  & 0.5 \\
        
        TOF difference requirement & 0.2 \\
        
        Photon veto & 1.1 \\
        
        $|\Delta\theta|$ and $|\Delta\phi|$ requirement & $3.8$ \\
        
        Bhabha veto & $3.3$ \\
        
        Signal region & $1.7$ \\
        
        MC model & 10.0 \\
        
        MC statistics & 0.4 \\
        
        Number of $\psi(3686)$ events & 0.5 \\
        
        \hline
        
        Total &  11.4   \\
        
        \hline \hline
    \end{tabular}
\end{center}
\label{tab:syst}
\end{table*}

%
%
\section{Results}

Eight candidate events are observed in the signal region from the full dataset, while six background events are expected. Therefore, no obvious excess is observed, and an UL on the BF $\mathcal{B}(\psi(3686)\to e^\pm\mu^\mp)$ is estimated using the profile likelihood method~\cite{ref:likelihood}. The likelihood function is defined as
\begin{equation}
\begin{aligned}
      \mathcal{L}(\mathcal{B}& ,\epsilon_{\rm sig}, N_{\rm \psi(3686)}, N_{\rm bkg1}, N_{\rm bkg2}) = \\
      &\mathcal{P}(N_{\rm obs}|N^{\rm data}_{\rm \psi(3686)}\cdot\mathcal{B}\cdot\epsilon_{\rm sig} + N_{\rm bkg1} + N_{\rm bkg2})\\
      &\cdot \mathcal{G} (\epsilon_{\rm sig}|\epsilon^{\rm MC}_{\rm sig}, \epsilon^{\rm MC}_{\rm sig}\cdot\sigma^{\rm MC}_{\rm sig}) \\
      &\cdot \mathcal{P} (N^{\rm \psi(3686)-MC}_{\rm bkg1}|N_{\rm bkg1}/f_1) \\
      &\cdot \prod_{k} \mathcal{P} (N^{k}_{\rm cont}|N_{\rm bkg2}/f^k_2) \\
      &\cdot \mathcal{G} (N_{\psi(3686)}, N^{\rm data}_{\psi(3686)}, \sigma_{N^{
      \rm data}_{\rm \psi(3686)}}),
\end{aligned}
\label{eq:likelihood}
\end{equation}
where the number of observed events, $N_{\rm obs}$, follows a Poisson distribution ($\mathcal{P}$) with an expected number equal to the sum of signal events ($N^{\rm data}_{\rm \psi(3686)}\cdot\mathcal{B}\cdot\epsilon_{\rm sig}$) and background events ($N_{\rm bkg1} + N_{\rm bkg2}$). Here, $N^{\rm data}_{\rm \psi(3686)}$ is the number of $\psi(3686)$ events in the data sample, $\mathcal{B}$ is the BF of $\psi(3686)\to e^{\pm}\mu^{\mp}$, $\epsilon_{\rm sig}$ is the signal detection efficiency, and $N_{\rm bkg1}$ and $N_{\rm bkg2}$ are the expected numbers of background events from $\psi(3686)$ decay and continuum production, respectively. The detection efficiency, $\epsilon_{\rm sig}$, follows a Gaussian distribution with a mean given by the efficiency, which is determined from the MC sample, $\epsilon^{\rm MC}_{\rm sig}$, and a standard deviation of $\epsilon^{\rm MC}_{\rm sig}\cdot\sigma^{\rm MC}_{\rm sig}$. Here, $\sigma^{\rm MC}_{\rm sig}$ represents the combined statistical and systematic uncertainty of the efficiency. The number of different background ($N^{\rm \psi(3686)-MC}_{\rm bkg1}$, $N^{k}_{\rm cont}$) follows a Poisson distribution with expected values of $N_{\rm bkg1}/f_1$ and $N_{\rm bkg2}/f_2$, respectively, where $f_1$ and $f_2$ are the scale factors for different background sources. The number of $\psi(3686)$ events, $N_{\psi(3686)}$, follows a Gaussian distribution with a mean of $N^{\rm data}_{\psi(3686)}$ and a standard deviation of $\sigma_{N^{\rm data}_{\psi(3686)}}$. The values of the parameters in Eq.~\eqref{eq:likelihood} are listed in Table~\ref{tab:value}.

\begingroup
\renewcommand\arraystretch{1.2}
\begin{table*}[!ht]
\setlength{\abovecaptionskip}{0.0cm}
\setlength{\belowcaptionskip}{0.3cm}
\caption{The values of the parameters for the likelihood function in Eq.~\eqref{eq:likelihood}.}
\centering
\footnotesize
\begin{tabular}{lc}
\hline \hline
                    Parameter & Value\\
                    \hline
                    $N_{\rm obs}$ & $8$ \\
                    $N^{\rm data}_{\rm \psi(3686)}$ & $2367.0\times10^{6}$\\
                    $\sigma_{N^{\rm data}_{\rm \psi(3686)}}$ & $11.1\times10^{6}$\\
                    $\epsilon^{\rm MC}_{\rm sig}$ & $24.18\%$\\
                    $\sigma^{\rm MC}_{\rm sig}$ & $11.4\%$\\
                    $N^{\psi(3686)-{\rm MC}}_{\rm bkg1}$ &
                    $18$\\
                    $N^{3.773}_{\rm cont}$ &$10$ \\
                    $N^{3.682}_{\rm cont}$ &$1$ \\
                    $N^{3.650}_{\rm cont}$ &$0$\\
                    $f_{1}$ & 0.0894\\
                    $f^{3.773}_{2}$ & 7.604\\
                    $f^{3.682}_{2}$ & 8.797\\
                    $f^{3.650}_{2}$ & 0.471\\
                
\hline \hline
\end{tabular}
\label{tab:value}
\end{table*}
\endgroup

\begin{figure}[htbp]
\begin{center}
\includegraphics[width=0.5\textwidth]{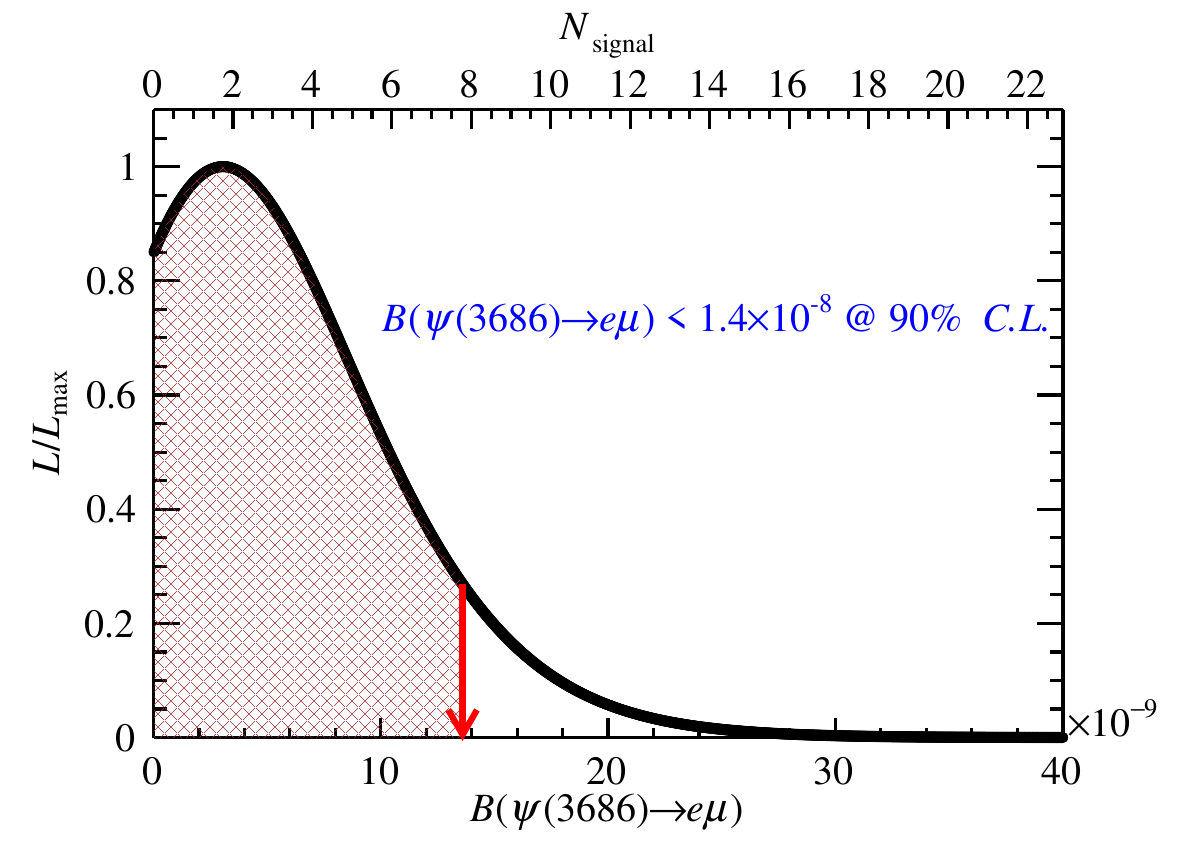}
\caption{
  The distribution of the normalized likelihood function versus BF for the CLFV decay $\psi(3686)\to e^\pm\mu^\mp$. $N_{\rm signal}$ is the number of signal events corresponding to the BF of $\psi(3686)\to e\mu$. The UL on its BF is set to be $\mathcal{B}(\psi(3686)\to e^{\pm}\mu^{\mp})<1.4\times 10^{-8}$ at the $90\%$ C.L.
  } \label{fig::likelihood}
\end{center}
\end{figure}

The UL of $\mathcal{B}(\psi(3686)\to e^{\pm}\mu^{\mp})$ is estimated by scanning the likelihood function of the BF. The scaled likelihood distribution, normalized by the maximum likelihood value $\mathcal{L}_{\rm max}$, is shown in Fig. \ref{fig::likelihood}. By integrating the likelihood function in the physical region where $\mathcal{B}\geq 0$, the UL on its BF is set to be $\mathcal{B}(\psi(3686)\to e^{\pm}\mu^{\mp})<1.4\times 10^{-8}$ at the $90\%$ C.L.

This result can be used to constrain the Wilson coefficients of effective operators in an effective field theory that accommodates CLFV~\cite{Hazard:2016fnc}. Combining the results of $J/\psi\to e^{\pm}\mu^{\mp}$~\cite{ref::jpsi2emu} and $\psi(3686)\to e^{\pm}\mu^{\mp}$, we get the most stringent constraints on the Wilson coefficients of $|C_{DL}^{e\mu}/\Lambda^2|~(|C_{DR}^{e\mu}/\Lambda^2|)$, $|C_{VL}^{ce\mu}/\Lambda^2|~(|C_{VR}^{ce\mu}/\Lambda^2|)$ and $|C_{TL}^{ce\mu}/\Lambda^2|~(|C_{TL}^{ce\mu}/\Lambda^2|)$, which are set to be $1.77\times10^{-4}~{\rm GeV^{-2}}$, $1.72\times10^{-6}~{\rm GeV^{-2}}$ and $7.75\times10^{-1}~{\rm GeV^{-2}}$, respectively.

%
%
\section{Summary}

In summary, based on $(2367.0\pm11.1)\times10^6$ $\psi(3686)$ events collected by the BESIII detector at the BEPCII collider, no significant excess of $\psi(3686)\to e^\pm\mu^\mp$ events is observed in the dataset compared to the expected background. The UL on its BF is determined to be $\mathcal{B}(\psi(3686)\to e^\pm\mu^\mp) < 1.4 \times 10^{-8}$ at the 90\% C.L. This is the first result in the search for CLFV in $\psi(3686)$ decays, which is helpful to constrain the parameters of new physics models, and provides new input for the effective field theory of CLFV~\cite{ref::psip2emu_significance}.

\section{Acknowledgments}

The BESIII Collaboration thanks the staff of BEPCII (https://cstr.cn/31109.02.BEPC) and the IHEP computing center for their strong support. This work is supported in part by National Key R\&D Program of China under Contracts Nos. 2023YFA1606000, 2023YFA1606704; National Natural Science Foundation of China (NSFC) under Contracts Nos. 11635010, 11935015, 11935016, 11935018, 12025502, 12035009, 12035013, 12061131003, 12192260, 12192261, 12192262, 12192263, 12192264, 12192265, 12221005, 12225509, 12235017, 12361141819; the Chinese Academy of Sciences (CAS) Large-Scale Scientific Facility Program; CAS under Contract No. YSBR-101; 100 Talents Program of CAS; The Institute of Nuclear and Particle Physics (INPAC) and Shanghai Key Laboratory for Particle Physics and Cosmology; Agencia Nacional de Investigación y Desarrollo de Chile (ANID), Chile under Contract No. ANID PIA/APOYO AFB230003; German Research Foundation DFG under Contract No. FOR5327; Istituto Nazionale di Fisica Nucleare, Italy; Knut and Alice Wallenberg Foundation under Contracts Nos. 2021.0174, 2021.0299; Ministry of Development of Turkey under Contract No. DPT2006K-120470; National Research Foundation of Korea under Contract No. NRF-2022R1A2C1092335; National Science and Technology fund of Mongolia; National Science Research and Innovation Fund (NSRF) via the Program Management Unit for Human Resources \& Institutional Development, Research and Innovation of Thailand under Contract No. B50G670107; Polish National Science Centre under Contract No. 2024/53/B/ST2/00975; Swedish Research Council under Contract No. 2019.04595; U. S. Department of Energy under Contract No. DE-FG02-05ER41374

\bibliographystyle{apsrev4-1}
\bibliography{./emu.bib}

\begin{thebibliography}{56}%
\makeatletter
\providecommand \@ifxundefined [1]{%
 \@ifx{#1\undefined}
}%
\providecommand \@ifnum [1]{%
 \ifnum #1\expandafter \@firstoftwo
 \else \expandafter \@secondoftwo
 \fi
}%
\providecommand \@ifx [1]{%
 \ifx #1\expandafter \@firstoftwo
 \else \expandafter \@secondoftwo
 \fi
}%
\providecommand \natexlab [1]{#1}%
\providecommand \enquote  [1]{``#1''}%
\providecommand \bibnamefont  [1]{#1}%
\providecommand \bibfnamefont [1]{#1}%
\providecommand \citenamefont [1]{#1}%
\providecommand \href@noop [0]{\@secondoftwo}%
\providecommand \href [0]{\begingroup \@sanitize@url \@href}%
\providecommand \@href[1]{\@@startlink{#1}\@@href}%
\providecommand \@@href[1]{\endgroup#1\@@endlink}%
\providecommand \@sanitize@url [0]{\catcode `\\12\catcode `\$12\catcode `\&12\catcode `\#12\catcode `\^12\catcode `\_12\catcode `\%12\relax}%
\providecommand \@@startlink[1]{}%
\providecommand \@@endlink[0]{}%
\providecommand \url  [0]{\begingroup\@sanitize@url \@url }%
\providecommand \@url [1]{\endgroup\@href {#1}{\urlprefix }}%
\providecommand \urlprefix  [0]{URL }%
\providecommand \Eprint [0]{\href }%
\providecommand \doibase [0]{http://dx.doi.org/}%
\providecommand \selectlanguage [0]{\@gobble}%
\providecommand \bibinfo  [0]{\@secondoftwo}%
\providecommand \bibfield  [0]{\@secondoftwo}%
\providecommand \translation [1]{[#1]}%
\providecommand \BibitemOpen [0]{}%
\providecommand \bibitemStop [0]{}%
\providecommand \bibitemNoStop [0]{.\EOS\space}%
\providecommand \EOS [0]{\spacefactor3000\relax}%
\providecommand \BibitemShut  [1]{\csname bibitem#1\endcsname}%
\let\auto@bib@innerbib\@empty
\bibitem [{\citenamefont {An}\ \emph {et~al.}(2012)\citenamefont {An} \emph {et~al.}}]{ref::neutrino_1}%
  \BibitemOpen
  \bibfield  {author} {\bibinfo {author} {\bibfnamefont {F.~P.}\ \bibnamefont {An}} \emph {et~al.} (\bibinfo {collaboration} {Daya Bay Collaboration}),\ }\href {\doibase 10.1103/PhysRevLett.108.171803} {\bibfield  {journal} {\bibinfo  {journal} {Phys. Rev. Lett.}\ }\textbf {\bibinfo {volume} {108}},\ \bibinfo {pages} {171803} (\bibinfo {year} {2012})},\ \Eprint {http://arxiv.org/abs/1203.1669} {arXiv:1203.1669 [hep-ex]} \BibitemShut {NoStop}%
\bibitem [{\citenamefont {Bernstein}\ and\ \citenamefont {Cooper}(2013)}]{ref::clfvreview_1}%
  \BibitemOpen
  \bibfield  {author} {\bibinfo {author} {\bibfnamefont {R.~H.}\ \bibnamefont {Bernstein}}\ and\ \bibinfo {author} {\bibfnamefont {P.~S.}\ \bibnamefont {Cooper}},\ }\href {\doibase 10.1016/j.physrep.2013.07.002} {\bibfield  {journal} {\bibinfo  {journal} {Phys. Rept.}\ }\textbf {\bibinfo {volume} {532}},\ \bibinfo {pages} {27} (\bibinfo {year} {2013})},\ \Eprint {http://arxiv.org/abs/1307.5787} {arXiv:1307.5787 [hep-ex]} \BibitemShut {NoStop}%
\bibitem [{\citenamefont {Ardu}\ and\ \citenamefont {Pezzullo}(2022)}]{ref::CLFVreview}%
  \BibitemOpen
  \bibfield  {author} {\bibinfo {author} {\bibfnamefont {M.}~\bibnamefont {Ardu}}\ and\ \bibinfo {author} {\bibfnamefont {G.}~\bibnamefont {Pezzullo}},\ }\href {\doibase 10.3390/universe8060299} {\bibfield  {journal} {\bibinfo  {journal} {Universe}\ }\textbf {\bibinfo {volume} {8}},\ \bibinfo {pages} {299} (\bibinfo {year} {2022})},\ \Eprint {http://arxiv.org/abs/2204.08220} {arXiv:2204.08220 [hep-ph]} \BibitemShut {NoStop}%
\bibitem [{\citenamefont {Afanaciev}\ \emph {et~al.}(2024)\citenamefont {Afanaciev} \emph {et~al.}}]{ref::mu2egamma}%
  \BibitemOpen
  \bibfield  {author} {\bibinfo {author} {\bibfnamefont {K.}~\bibnamefont {Afanaciev}} \emph {et~al.} (\bibinfo {collaboration} {MEG II Collaboration}),\ }\href {\doibase 10.1140/epjc/s10052-024-12416-2} {\bibfield  {journal} {\bibinfo  {journal} {Eur. Phys. J. C}\ }\textbf {\bibinfo {volume} {84}},\ \bibinfo {pages} {216} (\bibinfo {year} {2024})},\ \Eprint {http://arxiv.org/abs/2310.12614} {arXiv:2310.12614 [hep-ex]} \BibitemShut {NoStop}%
\bibitem [{\citenamefont {Bellgardt}\ \emph {et~al.}(1988)\citenamefont {Bellgardt} \emph {et~al.}}]{ref::mu2eee}%
  \BibitemOpen
  \bibfield  {author} {\bibinfo {author} {\bibfnamefont {U.}~\bibnamefont {Bellgardt}} \emph {et~al.} (\bibinfo {collaboration} {SINDRUM Collaboration}),\ }\href {\doibase 10.1016/0550-3213(88)90462-2} {\bibfield  {journal} {\bibinfo  {journal} {Nucl. Phys. B}\ }\textbf {\bibinfo {volume} {299}},\ \bibinfo {pages} {1} (\bibinfo {year} {1988})}\BibitemShut {NoStop}%
\bibitem [{\citenamefont {Bertl}\ \emph {et~al.}(2006)\citenamefont {Bertl} \emph {et~al.}}]{ref::muN2eN_1}%
  \BibitemOpen
  \bibfield  {author} {\bibinfo {author} {\bibfnamefont {W.~H.}\ \bibnamefont {Bertl}} \emph {et~al.} (\bibinfo {collaboration} {SINDRUM II Collaboration}),\ }\href {\doibase 10.1140/epjc/s2006-02582-x} {\bibfield  {journal} {\bibinfo  {journal} {Eur. Phys. J. C}\ }\textbf {\bibinfo {volume} {47}},\ \bibinfo {pages} {337} (\bibinfo {year} {2006})}\BibitemShut {NoStop}%
\bibitem [{\citenamefont {Dohmen}\ \emph {et~al.}(1993)\citenamefont {Dohmen} \emph {et~al.}}]{ref::muN2eN_2}%
  \BibitemOpen
  \bibfield  {author} {\bibinfo {author} {\bibfnamefont {C.}~\bibnamefont {Dohmen}} \emph {et~al.} (\bibinfo {collaboration} {SINDRUM II Collaboration}),\ }\href {\doibase 10.1016/0370-2693(93)91383-X} {\bibfield  {journal} {\bibinfo  {journal} {Phys. Lett. B}\ }\textbf {\bibinfo {volume} {317}},\ \bibinfo {pages} {631} (\bibinfo {year} {1993})}\BibitemShut {NoStop}%
\bibitem [{\citenamefont {Kaulard}\ \emph {et~al.}(1998)\citenamefont {Kaulard} \emph {et~al.}}]{ref::muN2eNprime}%
  \BibitemOpen
  \bibfield  {author} {\bibinfo {author} {\bibfnamefont {J.}~\bibnamefont {Kaulard}} \emph {et~al.} (\bibinfo {collaboration} {SINDRUM II Collaboration}),\ }\href {\doibase 10.1016/S0370-2693(97)01423-8} {\bibfield  {journal} {\bibinfo  {journal} {Phys. Lett. B}\ }\textbf {\bibinfo {volume} {422}},\ \bibinfo {pages} {334} (\bibinfo {year} {1998})}\BibitemShut {NoStop}%
\bibitem [{\citenamefont {Aubert}\ \emph {et~al.}(2010)\citenamefont {Aubert} \emph {et~al.}}]{ref::tau2lgamma}%
  \BibitemOpen
  \bibfield  {author} {\bibinfo {author} {\bibfnamefont {B.}~\bibnamefont {Aubert}} \emph {et~al.} (\bibinfo {collaboration} {BaBar Collaboration}),\ }\href {\doibase 10.1103/PhysRevLett.104.021802} {\bibfield  {journal} {\bibinfo  {journal} {Phys. Rev. Lett.}\ }\textbf {\bibinfo {volume} {104}},\ \bibinfo {pages} {021802} (\bibinfo {year} {2010})},\ \Eprint {http://arxiv.org/abs/0908.2381} {arXiv:0908.2381 [hep-ex]} \BibitemShut {NoStop}%
\bibitem [{\citenamefont {Hayasaka}\ \emph {et~al.}(2010)\citenamefont {Hayasaka} \emph {et~al.}}]{ref::tau2lll_1}%
  \BibitemOpen
  \bibfield  {author} {\bibinfo {author} {\bibfnamefont {K.}~\bibnamefont {Hayasaka}} \emph {et~al.},\ }\href {\doibase 10.1016/j.physletb.2010.03.037} {\bibfield  {journal} {\bibinfo  {journal} {Phys. Lett. B}\ }\textbf {\bibinfo {volume} {687}},\ \bibinfo {pages} {139} (\bibinfo {year} {2010})},\ \Eprint {http://arxiv.org/abs/1001.3221} {arXiv:1001.3221 [hep-ex]} \BibitemShut {NoStop}%
\bibitem [{\citenamefont {Sirunyan}\ \emph {et~al.}(2021)\citenamefont {Sirunyan} \emph {et~al.}}]{ref::tau2lll_2}%
  \BibitemOpen
  \bibfield  {author} {\bibinfo {author} {\bibfnamefont {A.~M.}\ \bibnamefont {Sirunyan}} \emph {et~al.} (\bibinfo {collaboration} {CMS Collaboration}),\ }\href {\doibase 10.1007/JHEP01(2021)163} {\bibfield  {journal} {\bibinfo  {journal} {JHEP}\ }\textbf {\bibinfo {volume} {01}},\ \bibinfo {pages} {163} (\bibinfo {year} {2021})},\ \Eprint {http://arxiv.org/abs/2007.05658} {arXiv:2007.05658 [hep-ex]} \BibitemShut {NoStop}%
\bibitem [{\citenamefont {Miyazaki}\ \emph {et~al.}(2011{\natexlab{a}})\citenamefont {Miyazaki} \emph {et~al.}}]{ref::tau2hl_1}%
  \BibitemOpen
  \bibfield  {author} {\bibinfo {author} {\bibfnamefont {Y.}~\bibnamefont {Miyazaki}} \emph {et~al.} (\bibinfo {collaboration} {Belle Collaboration}),\ }\href {\doibase 10.1016/j.physletb.2011.04.011} {\bibfield  {journal} {\bibinfo  {journal} {Phys. Lett. B}\ }\textbf {\bibinfo {volume} {699}},\ \bibinfo {pages} {251} (\bibinfo {year} {2011}{\natexlab{a}})},\ \Eprint {http://arxiv.org/abs/1101.0755} {arXiv:1101.0755 [hep-ex]} \BibitemShut {NoStop}%
\bibitem [{\citenamefont {Miyazaki}\ \emph {et~al.}(2011{\natexlab{b}})\citenamefont {Miyazaki} \emph {et~al.}}]{ref::tau2hl_2}%
  \BibitemOpen
  \bibfield  {author} {\bibinfo {author} {\bibfnamefont {Y.}~\bibnamefont {Miyazaki}} \emph {et~al.} (\bibinfo {collaboration} {Belle Collaboration}),\ }\href {\doibase 10.1016/j.physletb.2011.04.011} {\bibfield  {journal} {\bibinfo  {journal} {Phys. Lett. B}\ }\textbf {\bibinfo {volume} {699}},\ \bibinfo {pages} {251} (\bibinfo {year} {2011}{\natexlab{b}})},\ \Eprint {http://arxiv.org/abs/1101.0755} {arXiv:1101.0755 [hep-ex]} \BibitemShut {NoStop}%
\bibitem [{\citenamefont {Ablikim}\ \emph {et~al.}(2023)\citenamefont {Ablikim} \emph {et~al.}}]{ref::jpsi2emu}%
  \BibitemOpen
  \bibfield  {author} {\bibinfo {author} {\bibfnamefont {M.}~\bibnamefont {Ablikim}} \emph {et~al.} (\bibinfo {collaboration} {BESIII Collaboration}),\ }\href {\doibase 10.1007/s11433-022-1995-0} {\bibfield  {journal} {\bibinfo  {journal} {Sci. China Phys. Mech. Astron.}\ }\textbf {\bibinfo {volume} {66}},\ \bibinfo {pages} {221011} (\bibinfo {year} {2023})},\ \Eprint {http://arxiv.org/abs/2206.13956} {arXiv:2206.13956 [hep-ex]} \BibitemShut {NoStop}%
\bibitem [{\citenamefont {Ablikim}\ \emph {et~al.}(2021)\citenamefont {Ablikim} \emph {et~al.}}]{ref::jpsi2etau}%
  \BibitemOpen
  \bibfield  {author} {\bibinfo {author} {\bibfnamefont {M.}~\bibnamefont {Ablikim}} \emph {et~al.} (\bibinfo {collaboration} {BESIII Collaboration}),\ }\href {\doibase 10.1103/PhysRevD.103.112007} {\bibfield  {journal} {\bibinfo  {journal} {Phys. Rev. D}\ }\textbf {\bibinfo {volume} {103}},\ \bibinfo {pages} {112007} (\bibinfo {year} {2021})},\ \Eprint {http://arxiv.org/abs/2103.11540} {arXiv:2103.11540 [hep-ex]} \BibitemShut {NoStop}%
\bibitem [{\citenamefont {Patra}\ \emph {et~al.}(2022)\citenamefont {Patra} \emph {et~al.}}]{ref::upsilon2ll_1}%
  \BibitemOpen
  \bibfield  {author} {\bibinfo {author} {\bibfnamefont {S.}~\bibnamefont {Patra}} \emph {et~al.} (\bibinfo {collaboration} {Belle Collaboration}),\ }\href {\doibase 10.1007/JHEP05(2022)095} {\bibfield  {journal} {\bibinfo  {journal} {JHEP}\ }\textbf {\bibinfo {volume} {05}},\ \bibinfo {pages} {095} (\bibinfo {year} {2022})},\ \Eprint {http://arxiv.org/abs/2201.09620} {arXiv:2201.09620 [hep-ex]} \BibitemShut {NoStop}%
\bibitem [{\citenamefont {Lees}\ \emph {et~al.}(2022)\citenamefont {Lees} \emph {et~al.}}]{ref::upsilon2ll_2}%
  \BibitemOpen
  \bibfield  {author} {\bibinfo {author} {\bibfnamefont {J.~P.}\ \bibnamefont {Lees}} \emph {et~al.} (\bibinfo {collaboration} {BaBar Collaboration}),\ }\href {\doibase 10.1103/PhysRevLett.128.091804} {\bibfield  {journal} {\bibinfo  {journal} {Phys. Rev. Lett.}\ }\textbf {\bibinfo {volume} {128}},\ \bibinfo {pages} {091804} (\bibinfo {year} {2022})},\ \Eprint {http://arxiv.org/abs/2109.03364} {arXiv:2109.03364 [hep-ex]} \BibitemShut {NoStop}%
\bibitem [{\citenamefont {Hou}\ \emph {et~al.}(2022)\citenamefont {Hou}, \citenamefont {Kumar},\ and\ \citenamefont {Teunissen}}]{ref::yukawa}%
  \BibitemOpen
  \bibfield  {author} {\bibinfo {author} {\bibfnamefont {W.-S.}\ \bibnamefont {Hou}}, \bibinfo {author} {\bibfnamefont {G.}~\bibnamefont {Kumar}}, \ and\ \bibinfo {author} {\bibfnamefont {S.}~\bibnamefont {Teunissen}},\ }\href {\doibase 10.1007/JHEP01(2022)092} {\bibfield  {journal} {\bibinfo  {journal} {JHEP}\ }\textbf {\bibinfo {volume} {01}},\ \bibinfo {pages} {092} (\bibinfo {year} {2022})},\ \Eprint {http://arxiv.org/abs/2109.08936} {arXiv:2109.08936 [hep-ph]} \BibitemShut {NoStop}%
\bibitem [{\citenamefont {Escribano}\ \emph {et~al.}(2022)\citenamefont {Escribano}, \citenamefont {Hirsch}, \citenamefont {Nava},\ and\ \citenamefont {Vicente}}]{ref::majorana}%
  \BibitemOpen
  \bibfield  {author} {\bibinfo {author} {\bibfnamefont {P.}~\bibnamefont {Escribano}}, \bibinfo {author} {\bibfnamefont {M.}~\bibnamefont {Hirsch}}, \bibinfo {author} {\bibfnamefont {J.}~\bibnamefont {Nava}}, \ and\ \bibinfo {author} {\bibfnamefont {A.}~\bibnamefont {Vicente}},\ }\href {\doibase 10.1007/JHEP01(2022)098} {\bibfield  {journal} {\bibinfo  {journal} {JHEP}\ }\textbf {\bibinfo {volume} {01}},\ \bibinfo {pages} {098} (\bibinfo {year} {2022})},\ \Eprint {http://arxiv.org/abs/2108.01101} {arXiv:2108.01101 [hep-ph]} \BibitemShut {NoStop}%
\bibitem [{\citenamefont {Dimopoulos}\ and\ \citenamefont {Georgi}(1981)}]{ref::susy_gut}%
  \BibitemOpen
  \bibfield  {author} {\bibinfo {author} {\bibfnamefont {S.}~\bibnamefont {Dimopoulos}}\ and\ \bibinfo {author} {\bibfnamefont {H.}~\bibnamefont {Georgi}},\ }\href {\doibase 10.1016/0550-3213(81)90522-8} {\bibfield  {journal} {\bibinfo  {journal} {Nucl. Phys. B}\ }\textbf {\bibinfo {volume} {193}},\ \bibinfo {pages} {150} (\bibinfo {year} {1981})}\BibitemShut {NoStop}%
\bibitem [{\citenamefont {Kitano}\ and\ \citenamefont {Yamamoto}(2000)}]{ref::susy_leptons}%
  \BibitemOpen
  \bibfield  {author} {\bibinfo {author} {\bibfnamefont {R.}~\bibnamefont {Kitano}}\ and\ \bibinfo {author} {\bibfnamefont {K.}~\bibnamefont {Yamamoto}},\ }\href {\doibase 10.1103/PhysRevD.62.073007} {\bibfield  {journal} {\bibinfo  {journal} {Phys. Rev. D}\ }\textbf {\bibinfo {volume} {62}},\ \bibinfo {pages} {073007} (\bibinfo {year} {2000})},\ \Eprint {http://arxiv.org/abs/hep-ph/0003063} {arXiv:hep-ph/0003063} \BibitemShut {NoStop}%
\bibitem [{\citenamefont {Borzumati}\ and\ \citenamefont {Masiero}(1986)}]{ref::susy_neutrino}%
  \BibitemOpen
  \bibfield  {author} {\bibinfo {author} {\bibfnamefont {F.}~\bibnamefont {Borzumati}}\ and\ \bibinfo {author} {\bibfnamefont {A.}~\bibnamefont {Masiero}},\ }\href {\doibase 10.1103/PhysRevLett.57.961} {\bibfield  {journal} {\bibinfo  {journal} {Phys. Rev. Lett.}\ }\textbf {\bibinfo {volume} {57}},\ \bibinfo {pages} {961} (\bibinfo {year} {1986})}\BibitemShut {NoStop}%
\bibitem [{\citenamefont {Bernabeu}\ \emph {et~al.}(1993)\citenamefont {Bernabeu}, \citenamefont {Nardi},\ and\ \citenamefont {Tommasini}}]{ref::susy_zprime}%
  \BibitemOpen
  \bibfield  {author} {\bibinfo {author} {\bibfnamefont {J.}~\bibnamefont {Bernabeu}}, \bibinfo {author} {\bibfnamefont {E.}~\bibnamefont {Nardi}}, \ and\ \bibinfo {author} {\bibfnamefont {D.}~\bibnamefont {Tommasini}},\ }\href {\doibase 10.1016/0550-3213(93)90446-V} {\bibfield  {journal} {\bibinfo  {journal} {Nucl. Phys. B}\ }\textbf {\bibinfo {volume} {409}},\ \bibinfo {pages} {69} (\bibinfo {year} {1993})},\ \Eprint {http://arxiv.org/abs/hep-ph/9306251} {arXiv:hep-ph/9306251} \BibitemShut {NoStop}%
\bibitem [{\citenamefont {Huang}\ \emph {et~al.}(2018)\citenamefont {Huang}, \citenamefont {P{\"a}s},\ and\ \citenamefont {Zeissner}}]{ref::clfvpsip_gut}%
  \BibitemOpen
  \bibfield  {author} {\bibinfo {author} {\bibfnamefont {W.-C.}\ \bibnamefont {Huang}}, \bibinfo {author} {\bibfnamefont {H.}~\bibnamefont {P{\"a}s}}, \ and\ \bibinfo {author} {\bibfnamefont {S.}~\bibnamefont {Zeissner}},\ }\href {\doibase 10.1103/PhysRevD.97.055040} {\bibfield  {journal} {\bibinfo  {journal} {Phys. Rev. D}\ }\textbf {\bibinfo {volume} {97}},\ \bibinfo {pages} {055040} (\bibinfo {year} {2018})},\ \Eprint {http://arxiv.org/abs/1608.04354} {arXiv:1608.04354 [hep-ph]} \BibitemShut {NoStop}%
\bibitem [{\citenamefont {Chang}\ and\ \citenamefont {Feng}(2000)}]{ref::clfvpsip_susy}%
  \BibitemOpen
  \bibfield  {author} {\bibinfo {author} {\bibfnamefont {C.-H.}\ \bibnamefont {Chang}}\ and\ \bibinfo {author} {\bibfnamefont {T.-F.}\ \bibnamefont {Feng}},\ }\href {\doibase 10.1007/s100529900194} {\bibfield  {journal} {\bibinfo  {journal} {Eur. Phys. J. C}\ }\textbf {\bibinfo {volume} {12}},\ \bibinfo {pages} {137} (\bibinfo {year} {2000})},\ \Eprint {http://arxiv.org/abs/hep-ph/9901260} {arXiv:hep-ph/9901260} \BibitemShut {NoStop}%
\bibitem [{\citenamefont {Cvetic}(1999)}]{ref::clfvpsip_tc2}%
  \BibitemOpen
  \bibfield  {author} {\bibinfo {author} {\bibfnamefont {G.}~\bibnamefont {Cvetic}},\ }\href {\doibase 10.1103/RevModPhys.71.513} {\bibfield  {journal} {\bibinfo  {journal} {Rev. Mod. Phys.}\ }\textbf {\bibinfo {volume} {71}},\ \bibinfo {pages} {513} (\bibinfo {year} {1999})},\ \Eprint {http://arxiv.org/abs/hep-ph/9702381} {arXiv:hep-ph/9702381} \BibitemShut {NoStop}%
\bibitem [{\citenamefont {Nussinov}\ \emph {et~al.}(2001)\citenamefont {Nussinov}, \citenamefont {Peccei},\ and\ \citenamefont {Zhang}}]{ref::theory_clfvpsip_1}%
  \BibitemOpen
  \bibfield  {author} {\bibinfo {author} {\bibfnamefont {S.}~\bibnamefont {Nussinov}}, \bibinfo {author} {\bibfnamefont {R.~D.}\ \bibnamefont {Peccei}}, \ and\ \bibinfo {author} {\bibfnamefont {X.~M.}\ \bibnamefont {Zhang}},\ }\href {\doibase 10.1103/PhysRevD.63.016003} {\bibfield  {journal} {\bibinfo  {journal} {Phys. Rev. D}\ }\textbf {\bibinfo {volume} {63}},\ \bibinfo {pages} {016003} (\bibinfo {year} {2001})},\ \Eprint {http://arxiv.org/abs/hep-ph/0004153} {arXiv:hep-ph/0004153} \BibitemShut {NoStop}%
\bibitem [{\citenamefont {Gutsche}\ \emph {et~al.}(2011)\citenamefont {Gutsche}, \citenamefont {Helo}, \citenamefont {Kovalenko},\ and\ \citenamefont {Lyubovitskij}}]{ref::theory_clfvpsip_2}%
  \BibitemOpen
  \bibfield  {author} {\bibinfo {author} {\bibfnamefont {T.}~\bibnamefont {Gutsche}}, \bibinfo {author} {\bibfnamefont {J.~C.}\ \bibnamefont {Helo}}, \bibinfo {author} {\bibfnamefont {S.}~\bibnamefont {Kovalenko}}, \ and\ \bibinfo {author} {\bibfnamefont {V.~E.}\ \bibnamefont {Lyubovitskij}},\ }\href {\doibase 10.1103/PhysRevD.83.115015} {\bibfield  {journal} {\bibinfo  {journal} {Phys. Rev. D}\ }\textbf {\bibinfo {volume} {83}},\ \bibinfo {pages} {115015} (\bibinfo {year} {2011})},\ \Eprint {http://arxiv.org/abs/1103.1317} {arXiv:1103.1317 [hep-ph]} \BibitemShut {NoStop}%
\bibitem [{\citenamefont {Bordes}\ \emph {et~al.}(2001)\citenamefont {Bordes}, \citenamefont {Chan},\ and\ \citenamefont {Tsou}}]{ref::theory_clfvpsip_3}%
  \BibitemOpen
  \bibfield  {author} {\bibinfo {author} {\bibfnamefont {J.}~\bibnamefont {Bordes}}, \bibinfo {author} {\bibfnamefont {H.-M.}\ \bibnamefont {Chan}}, \ and\ \bibinfo {author} {\bibfnamefont {S.~T.}\ \bibnamefont {Tsou}},\ }\href {\doibase 10.1103/PhysRevD.63.016006} {\bibfield  {journal} {\bibinfo  {journal} {Phys. Rev. D}\ }\textbf {\bibinfo {volume} {63}},\ \bibinfo {pages} {016006} (\bibinfo {year} {2001})},\ \Eprint {http://arxiv.org/abs/hep-ph/0006338} {arXiv:hep-ph/0006338} \BibitemShut {NoStop}%
\bibitem [{\citenamefont {Sun}\ \emph {et~al.}(2012)\citenamefont {Sun}, \citenamefont {Feng}, \citenamefont {Kou}, \citenamefont {Sun}, \citenamefont {Gao},\ and\ \citenamefont {Zhang}}]{ref::theory_clfvpsip_4}%
  \BibitemOpen
  \bibfield  {author} {\bibinfo {author} {\bibfnamefont {K.-S.}\ \bibnamefont {Sun}}, \bibinfo {author} {\bibfnamefont {T.-F.}\ \bibnamefont {Feng}}, \bibinfo {author} {\bibfnamefont {L.-N.}\ \bibnamefont {Kou}}, \bibinfo {author} {\bibfnamefont {F.}~\bibnamefont {Sun}}, \bibinfo {author} {\bibfnamefont {T.-J.}\ \bibnamefont {Gao}}, \ and\ \bibinfo {author} {\bibfnamefont {H.-B.}\ \bibnamefont {Zhang}},\ }\href {\doibase 10.1142/S0217732312501726} {\bibfield  {journal} {\bibinfo  {journal} {Mod. Phys. Lett. A}\ }\textbf {\bibinfo {volume} {27}},\ \bibinfo {pages} {1250172} (\bibinfo {year} {2012})},\ \Eprint {http://arxiv.org/abs/1312.2072} {arXiv:1312.2072 [hep-ph]} \BibitemShut {NoStop}%
\bibitem [{\citenamefont {Hazard}\ and\ \citenamefont {Petrov}(2016{\natexlab{a}})}]{ref::theory_clfvpsip_5}%
  \BibitemOpen
  \bibfield  {author} {\bibinfo {author} {\bibfnamefont {D.~E.}\ \bibnamefont {Hazard}}\ and\ \bibinfo {author} {\bibfnamefont {A.~A.}\ \bibnamefont {Petrov}},\ }\href {\doibase 10.1103/PhysRevD.94.074023} {\bibfield  {journal} {\bibinfo  {journal} {Phys. Rev. D}\ }\textbf {\bibinfo {volume} {94}},\ \bibinfo {pages} {074023} (\bibinfo {year} {2016}{\natexlab{a}})},\ \Eprint {http://arxiv.org/abs/1607.00815} {arXiv:1607.00815 [hep-ph]} \BibitemShut {NoStop}%
\bibitem [{\citenamefont {Dong}\ \emph {et~al.}(2018)\citenamefont {Dong}, \citenamefont {Zhao}, \citenamefont {Feng}, \citenamefont {Ning}, \citenamefont {Chen}, \citenamefont {Zhang},\ and\ \citenamefont {Feng}}]{ref::theory_clfvpsip_6}%
  \BibitemOpen
  \bibfield  {author} {\bibinfo {author} {\bibfnamefont {X.-X.}\ \bibnamefont {Dong}}, \bibinfo {author} {\bibfnamefont {S.-M.}\ \bibnamefont {Zhao}}, \bibinfo {author} {\bibfnamefont {J.-J.}\ \bibnamefont {Feng}}, \bibinfo {author} {\bibfnamefont {G.-Z.}\ \bibnamefont {Ning}}, \bibinfo {author} {\bibfnamefont {J.-B.}\ \bibnamefont {Chen}}, \bibinfo {author} {\bibfnamefont {H.-B.}\ \bibnamefont {Zhang}}, \ and\ \bibinfo {author} {\bibfnamefont {T.-F.}\ \bibnamefont {Feng}},\ }\href {\doibase 10.1103/PhysRevD.97.056027} {\bibfield  {journal} {\bibinfo  {journal} {Phys. Rev. D}\ }\textbf {\bibinfo {volume} {97}},\ \bibinfo {pages} {056027} (\bibinfo {year} {2018})},\ \Eprint {http://arxiv.org/abs/1710.07408} {arXiv:1710.07408 [hep-ph]} \BibitemShut {NoStop}%
\bibitem [{\citenamefont {Ablikim}\ \emph {et~al.}(2024{\natexlab{a}})\citenamefont {Ablikim} \emph {et~al.}}]{ref:numpsip}%
  \BibitemOpen
  \bibfield  {author} {\bibinfo {author} {\bibfnamefont {M.}~\bibnamefont {Ablikim}} \emph {et~al.} (\bibinfo {collaboration} {BESIII Collaboration}),\ }\href {\doibase 10.1088/1674-1137/ad595b} {\bibfield  {journal} {\bibinfo  {journal} {Chin. Phys. C}\ }\textbf {\bibinfo {volume} {48}},\ \bibinfo {pages} {093001} (\bibinfo {year} {2024}{\natexlab{a}})},\ \Eprint {http://arxiv.org/abs/2403.06766} {arXiv:2403.06766 [hep-ex]} \BibitemShut {NoStop}%
\bibitem [{\citenamefont {Ablikim}(2013)}]{ref:data-3650-3773}%
  \BibitemOpen
  \bibfield  {author} {\bibinfo {author} {\bibfnamefont {M.}~\bibnamefont {Ablikim}} (\bibinfo {collaboration} {BESIII Collaboration}),\ }\href {\doibase 10.1088/1674-1137/37/12/123001} {\bibfield  {journal} {\bibinfo  {journal} {Chin. Phys. C}\ }\textbf {\bibinfo {volume} {37}},\ \bibinfo {pages} {123001} (\bibinfo {year} {2013})},\ \Eprint {http://arxiv.org/abs/1307.2022} {arXiv:1307.2022 [hep-ex]} \BibitemShut {NoStop}%
\bibitem [{\citenamefont {Ablikim}\ \emph {et~al.}(2024{\natexlab{b}})\citenamefont {Ablikim} \emph {et~al.}}]{ref:data-3773}%
  \BibitemOpen
  \bibfield  {author} {\bibinfo {author} {\bibfnamefont {M.}~\bibnamefont {Ablikim}} \emph {et~al.} (\bibinfo {collaboration} {BESIII Collaboration}),\ }\href {\doibase 10.1088/1674-1137/ad70a0} {\bibfield  {journal} {\bibinfo  {journal} {Chin. Phys. C}\ }\textbf {\bibinfo {volume} {48}},\ \bibinfo {pages} {123001} (\bibinfo {year} {2024}{\natexlab{b}})},\ \Eprint {http://arxiv.org/abs/2406.05827} {arXiv:2406.05827 [hep-ex]} \BibitemShut {NoStop}%
\bibitem [{\citenamefont {Liao}\ \emph {et~al.}(2025)\citenamefont {Liao}, \citenamefont {Liu}, \citenamefont {Wang}, \citenamefont {Sun},\ and\ \citenamefont {You}}]{Liao:2025lth}%
  \BibitemOpen
  \bibfield  {author} {\bibinfo {author} {\bibfnamefont {M.-H.}\ \bibnamefont {Liao}}, \bibinfo {author} {\bibfnamefont {J.-S.}\ \bibnamefont {Liu}}, \bibinfo {author} {\bibfnamefont {X.-N.}\ \bibnamefont {Wang}}, \bibinfo {author} {\bibfnamefont {S.-S.}\ \bibnamefont {Sun}}, \ and\ \bibinfo {author} {\bibfnamefont {Z.-Y.}\ \bibnamefont {You}},\ }\href {\doibase 10.1007/s41365-025-01789-y} {\bibfield  {journal} {\bibinfo  {journal} {Nucl. Sci. Tech.}\ }\textbf {\bibinfo {volume} {36}},\ \bibinfo {pages} {218} (\bibinfo {year} {2025})},\ \Eprint {http://arxiv.org/abs/2509.16066} {arXiv:2509.16066 [hep-ex]} \BibitemShut {NoStop}%
\bibitem [{\citenamefont {Ablikim}\ \emph {et~al.}(2010)\citenamefont {Ablikim} \emph {et~al.}}]{Ablikim:2009aa}%
  \BibitemOpen
  \bibfield  {author} {\bibinfo {author} {\bibfnamefont {M.}~\bibnamefont {Ablikim}} \emph {et~al.} (\bibinfo {collaboration} {BESIII Collaboration}),\ }\href {\doibase 10.1016/j.nima.2009.12.050} {\bibfield  {journal} {\bibinfo  {journal} {Nucl. Instrum. Meth. A}\ }\textbf {\bibinfo {volume} {614}},\ \bibinfo {pages} {345} (\bibinfo {year} {2010})},\ \Eprint {http://arxiv.org/abs/0911.4960} {arXiv:0911.4960 [physics.ins-det]} \BibitemShut {NoStop}%
\bibitem [{\citenamefont {Yu}\ \emph {et~al.}(2016)\citenamefont {Yu} \emph {et~al.}}]{Yu:IPAC2016-TUYA01}%
  \BibitemOpen
  \bibfield  {author} {\bibinfo {author} {\bibfnamefont {C.}~\bibnamefont {Yu}} \emph {et~al.},\ }in\ \href {\doibase 10.18429/JACoW-IPAC2016-TUYA01} {\emph {\bibinfo {booktitle} {{7th International Particle Accelerator Conference}}}}\ (\bibinfo {year} {2016})\ p.\ \bibinfo {pages} {TUYA01}\BibitemShut {NoStop}%
\bibitem [{\citenamefont {Lu}\ \emph {et~al.}(2020)\citenamefont {Lu}, \citenamefont {Xiao},\ and\ \citenamefont {Ji}}]{EcmsMea}%
  \BibitemOpen
  \bibfield  {author} {\bibinfo {author} {\bibfnamefont {J.}~\bibnamefont {Lu}}, \bibinfo {author} {\bibfnamefont {Y.}~\bibnamefont {Xiao}}, \ and\ \bibinfo {author} {\bibfnamefont {X.}~\bibnamefont {Ji}},\ }\href {\doibase 10.1007/s41605-020-00188-8} {\bibfield  {journal} {\bibinfo  {journal} {Radiat. Detect. Technol. Methods}\ }\textbf {\bibinfo {volume} {4}},\ \bibinfo {pages} {337} (\bibinfo {year} {2020})}\BibitemShut {NoStop}%
\bibitem [{\citenamefont {Li}\ \emph {et~al.}(2017)\citenamefont {Li} \emph {et~al.}}]{ref:tofupgrade}%
  \BibitemOpen
  \bibfield  {author} {\bibinfo {author} {\bibfnamefont {X.}~\bibnamefont {Li}} \emph {et~al.},\ }\href {\doibase 10.1007/s41605-017-0014-2} {\bibfield  {journal} {\bibinfo  {journal} {Radiat. Detect. Technol. Methods}\ }\textbf {\bibinfo {volume} {1}},\ \bibinfo {pages} {13} (\bibinfo {year} {2017})}\BibitemShut {NoStop}%
\bibitem [{\citenamefont {Agostinelli}\ \emph {et~al.}(2003)\citenamefont {Agostinelli} \emph {et~al.}}]{geant4}%
  \BibitemOpen
  \bibfield  {author} {\bibinfo {author} {\bibfnamefont {S.}~\bibnamefont {Agostinelli}} \emph {et~al.} (\bibinfo {collaboration} {GEANT4 Collaboration}),\ }\href {\doibase 10.1016/S0168-9002(03)01368-8} {\bibfield  {journal} {\bibinfo  {journal} {Nucl. Instrum. Meth. A}\ }\textbf {\bibinfo {volume} {506}},\ \bibinfo {pages} {250} (\bibinfo {year} {2003})}\BibitemShut {NoStop}%
\bibitem [{\citenamefont {Huang}\ \emph {et~al.}(2022)\citenamefont {Huang}, \citenamefont {Li}, \citenamefont {Qian}, \citenamefont {Zhu}, \citenamefont {Li}, \citenamefont {Zhang}, \citenamefont {Sun},\ and\ \citenamefont {You}}]{Huang:2022wuo}%
  \BibitemOpen
  \bibfield  {author} {\bibinfo {author} {\bibfnamefont {K.-X.}\ \bibnamefont {Huang}}, \bibinfo {author} {\bibfnamefont {Z.-J.}\ \bibnamefont {Li}}, \bibinfo {author} {\bibfnamefont {Z.}~\bibnamefont {Qian}}, \bibinfo {author} {\bibfnamefont {J.}~\bibnamefont {Zhu}}, \bibinfo {author} {\bibfnamefont {H.-Y.}\ \bibnamefont {Li}}, \bibinfo {author} {\bibfnamefont {Y.-M.}\ \bibnamefont {Zhang}}, \bibinfo {author} {\bibfnamefont {S.-S.}\ \bibnamefont {Sun}}, \ and\ \bibinfo {author} {\bibfnamefont {Z.-Y.}\ \bibnamefont {You}},\ }\href {\doibase 10.1007/s41365-022-01133-8} {\bibfield  {journal} {\bibinfo  {journal} {Nucl. Sci. Tech.}\ }\textbf {\bibinfo {volume} {33}},\ \bibinfo {pages} {142} (\bibinfo {year} {2022})},\ \Eprint {http://arxiv.org/abs/2206.10117} {arXiv:2206.10117 [physics.ins-det]} \BibitemShut {NoStop}%
\bibitem [{\citenamefont {Song}\ \emph {et~al.}(2026)\citenamefont {Song}, \citenamefont {Huang}, \citenamefont {Zeng}, \citenamefont {Liao}, \citenamefont {Wang}, \citenamefont {Zhang},\ and\ \citenamefont {You}}]{Song:2025pnt}%
  \BibitemOpen
  \bibfield  {author} {\bibinfo {author} {\bibfnamefont {T.-Z.}\ \bibnamefont {Song}}, \bibinfo {author} {\bibfnamefont {K.-X.}\ \bibnamefont {Huang}}, \bibinfo {author} {\bibfnamefont {Y.-J.}\ \bibnamefont {Zeng}}, \bibinfo {author} {\bibfnamefont {M.-H.}\ \bibnamefont {Liao}}, \bibinfo {author} {\bibfnamefont {X.-S.}\ \bibnamefont {Wang}}, \bibinfo {author} {\bibfnamefont {Y.-M.}\ \bibnamefont {Zhang}}, \ and\ \bibinfo {author} {\bibfnamefont {Z.-Y.}\ \bibnamefont {You}},\ }\href {\doibase 10.15302/frontphys.2026.026201} {\bibfield  {journal} {\bibinfo  {journal} {Front. Phys. (Beijing)}\ }\textbf {\bibinfo {volume} {21}},\ \bibinfo {pages} {26201} (\bibinfo {year} {2026})},\ \Eprint {http://arxiv.org/abs/2507.10261} {arXiv:2507.10261 [hep-ex]} \BibitemShut {NoStop}%
\bibitem [{\citenamefont {Jadach}\ \emph {et~al.}(2001)\citenamefont {Jadach}, \citenamefont {Ward},\ and\ \citenamefont {Was}}]{ref:kkmc1}%
  \BibitemOpen
  \bibfield  {author} {\bibinfo {author} {\bibfnamefont {S.}~\bibnamefont {Jadach}}, \bibinfo {author} {\bibfnamefont {B.~F.~L.}\ \bibnamefont {Ward}}, \ and\ \bibinfo {author} {\bibfnamefont {Z.}~\bibnamefont {Was}},\ }\href {\doibase 10.1103/PhysRevD.63.113009} {\bibfield  {journal} {\bibinfo  {journal} {Phys. Rev. D}\ }\textbf {\bibinfo {volume} {63}},\ \bibinfo {pages} {113009} (\bibinfo {year} {2001})},\ \Eprint {http://arxiv.org/abs/hep-ph/0006359} {arXiv:hep-ph/0006359} \BibitemShut {NoStop}%
\bibitem [{\citenamefont {Jadach}\ \emph {et~al.}(2000)\citenamefont {Jadach}, \citenamefont {Ward},\ and\ \citenamefont {Was}}]{ref:kkmc2}%
  \BibitemOpen
  \bibfield  {author} {\bibinfo {author} {\bibfnamefont {S.}~\bibnamefont {Jadach}}, \bibinfo {author} {\bibfnamefont {B.~F.~L.}\ \bibnamefont {Ward}}, \ and\ \bibinfo {author} {\bibfnamefont {Z.}~\bibnamefont {Was}},\ }\href {\doibase 10.1016/S0010-4655(00)00048-5} {\bibfield  {journal} {\bibinfo  {journal} {Comput. Phys. Commun.}\ }\textbf {\bibinfo {volume} {130}},\ \bibinfo {pages} {260} (\bibinfo {year} {2000})},\ \Eprint {http://arxiv.org/abs/hep-ph/9912214} {arXiv:hep-ph/9912214} \BibitemShut {NoStop}%
\bibitem [{\citenamefont {Lange}(2001)}]{ref:evtgen1}%
  \BibitemOpen
  \bibfield  {author} {\bibinfo {author} {\bibfnamefont {D.~J.}\ \bibnamefont {Lange}},\ }\href {\doibase 10.1016/S0168-9002(01)00089-4} {\bibfield  {journal} {\bibinfo  {journal} {Nucl. Instrum. Meth. A}\ }\textbf {\bibinfo {volume} {462}},\ \bibinfo {pages} {152} (\bibinfo {year} {2001})}\BibitemShut {NoStop}%
\bibitem [{\citenamefont {Ping}(2008)}]{ref:evtgen2}%
  \BibitemOpen
  \bibfield  {author} {\bibinfo {author} {\bibfnamefont {R.-G.}\ \bibnamefont {Ping}},\ }\href {\doibase 10.1088/1674-1137/32/8/001} {\bibfield  {journal} {\bibinfo  {journal} {Chin. Phys. C}\ }\textbf {\bibinfo {volume} {32}},\ \bibinfo {pages} {599} (\bibinfo {year} {2008})}\BibitemShut {NoStop}%
\bibitem [{\citenamefont {Navas}\ \emph {et~al.}(2024)\citenamefont {Navas} \emph {et~al.}}]{ref:pdg}%
  \BibitemOpen
  \bibfield  {author} {\bibinfo {author} {\bibfnamefont {S.}~\bibnamefont {Navas}} \emph {et~al.} (\bibinfo {collaboration} {Particle Data Group}),\ }\href {\doibase 10.1103/PhysRevD.110.030001} {\bibfield  {journal} {\bibinfo  {journal} {Phys. Rev. D}\ }\textbf {\bibinfo {volume} {110}},\ \bibinfo {pages} {030001} (\bibinfo {year} {2024})}\BibitemShut {NoStop}%
\bibitem [{\citenamefont {Chen}\ \emph {et~al.}(2000)\citenamefont {Chen}, \citenamefont {Huang}, \citenamefont {Qi}, \citenamefont {Zhang},\ and\ \citenamefont {Zhu}}]{ref:lundcharm1}%
  \BibitemOpen
  \bibfield  {author} {\bibinfo {author} {\bibfnamefont {J.~C.}\ \bibnamefont {Chen}}, \bibinfo {author} {\bibfnamefont {G.~S.}\ \bibnamefont {Huang}}, \bibinfo {author} {\bibfnamefont {X.~R.}\ \bibnamefont {Qi}}, \bibinfo {author} {\bibfnamefont {D.~H.}\ \bibnamefont {Zhang}}, \ and\ \bibinfo {author} {\bibfnamefont {Y.~S.}\ \bibnamefont {Zhu}},\ }\href {\doibase 10.1103/PhysRevD.62.034003} {\bibfield  {journal} {\bibinfo  {journal} {Phys. Rev. D}\ }\textbf {\bibinfo {volume} {62}},\ \bibinfo {pages} {034003} (\bibinfo {year} {2000})}\BibitemShut {NoStop}%
\bibitem [{\citenamefont {Yang}\ \emph {et~al.}(2014)\citenamefont {Yang}, \citenamefont {Ping},\ and\ \citenamefont {Chen}}]{ref:lundcharm2}%
  \BibitemOpen
  \bibfield  {author} {\bibinfo {author} {\bibfnamefont {R.-L.}\ \bibnamefont {Yang}}, \bibinfo {author} {\bibfnamefont {R.-G.}\ \bibnamefont {Ping}}, \ and\ \bibinfo {author} {\bibfnamefont {H.}~\bibnamefont {Chen}},\ }\href {\doibase 10.1088/0256-307X/31/6/061301} {\bibfield  {journal} {\bibinfo  {journal} {Chin. Phys. Lett.}\ }\textbf {\bibinfo {volume} {31}},\ \bibinfo {pages} {061301} (\bibinfo {year} {2014})}\BibitemShut {NoStop}%
\bibitem [{\citenamefont {Barberio}\ \emph {et~al.}(1991)\citenamefont {Barberio}, \citenamefont {van Eijk},\ and\ \citenamefont {Was}}]{photos2}%
  \BibitemOpen
  \bibfield  {author} {\bibinfo {author} {\bibfnamefont {E.}~\bibnamefont {Barberio}}, \bibinfo {author} {\bibfnamefont {B.}~\bibnamefont {van Eijk}}, \ and\ \bibinfo {author} {\bibfnamefont {Z.}~\bibnamefont {Was}},\ }\href {\doibase 10.1016/0010-4655(91)90012-A} {\bibfield  {journal} {\bibinfo  {journal} {Comput. Phys. Commun.}\ }\textbf {\bibinfo {volume} {66}},\ \bibinfo {pages} {115} (\bibinfo {year} {1991})}\BibitemShut {NoStop}%
\bibitem [{\citenamefont {Li}\ \emph {et~al.}(2024)\citenamefont {Li}, \citenamefont {Yuan}, \citenamefont {Song}, \citenamefont {Li}, \citenamefont {Li}, \citenamefont {Sun}, \citenamefont {Wang}, \citenamefont {You},\ and\ \citenamefont {Mao}}]{ref:visualbes3}%
  \BibitemOpen
  \bibfield  {author} {\bibinfo {author} {\bibfnamefont {Z.-J.}\ \bibnamefont {Li}}, \bibinfo {author} {\bibfnamefont {M.-K.}\ \bibnamefont {Yuan}}, \bibinfo {author} {\bibfnamefont {Y.-X.}\ \bibnamefont {Song}}, \bibinfo {author} {\bibfnamefont {Y.-G.}\ \bibnamefont {Li}}, \bibinfo {author} {\bibfnamefont {J.-S.}\ \bibnamefont {Li}}, \bibinfo {author} {\bibfnamefont {S.-S.}\ \bibnamefont {Sun}}, \bibinfo {author} {\bibfnamefont {X.-L.}\ \bibnamefont {Wang}}, \bibinfo {author} {\bibfnamefont {Z.-Y.}\ \bibnamefont {You}}, \ and\ \bibinfo {author} {\bibfnamefont {Y.-J.}\ \bibnamefont {Mao}},\ }\href {\doibase 10.1007/s11467-024-1422-7} {\bibfield  {journal} {\bibinfo  {journal} {Front. Phys. (Beijing)}\ }\textbf {\bibinfo {volume} {19}},\ \bibinfo {pages} {64201} (\bibinfo {year} {2024})},\ \Eprint {http://arxiv.org/abs/2404.07951} {arXiv:2404.07951 [physics.data-an]} \BibitemShut {NoStop}%
\bibitem [{\citenamefont {Punzi}(2003)}]{ref:punzi}%
  \BibitemOpen
  \bibfield  {author} {\bibinfo {author} {\bibfnamefont {G.}~\bibnamefont {Punzi}},\ }\href@noop {} {\bibfield  {journal} {\bibinfo  {journal} {eConf}\ }\textbf {\bibinfo {volume} {C030908}},\ \bibinfo {pages} {MODT002} (\bibinfo {year} {2003})},\ \Eprint {http://arxiv.org/abs/physics/0308063} {arXiv:physics/0308063} \BibitemShut {NoStop}%
\bibitem [{\citenamefont {Rolke}\ \emph {et~al.}(2005)\citenamefont {Rolke}, \citenamefont {Lopez},\ and\ \citenamefont {Conrad}}]{ref:likelihood}%
  \BibitemOpen
  \bibfield  {author} {\bibinfo {author} {\bibfnamefont {W.~A.}\ \bibnamefont {Rolke}}, \bibinfo {author} {\bibfnamefont {A.~M.}\ \bibnamefont {Lopez}}, \ and\ \bibinfo {author} {\bibfnamefont {J.}~\bibnamefont {Conrad}},\ }\href {\doibase 10.1016/j.nima.2005.05.068} {\bibfield  {journal} {\bibinfo  {journal} {Nucl. Instrum. Meth. A}\ }\textbf {\bibinfo {volume} {551}},\ \bibinfo {pages} {493} (\bibinfo {year} {2005})},\ \Eprint {http://arxiv.org/abs/physics/0403059} {arXiv:physics/0403059} \BibitemShut {NoStop}%
\bibitem [{\citenamefont {Hazard}\ and\ \citenamefont {Petrov}(2016{\natexlab{b}})}]{Hazard:2016fnc}%
  \BibitemOpen
  \bibfield  {author} {\bibinfo {author} {\bibfnamefont {D.~E.}\ \bibnamefont {Hazard}}\ and\ \bibinfo {author} {\bibfnamefont {A.~A.}\ \bibnamefont {Petrov}},\ }\href {\doibase 10.1103/PhysRevD.94.074023} {\bibfield  {journal} {\bibinfo  {journal} {Phys. Rev. D}\ }\textbf {\bibinfo {volume} {94}},\ \bibinfo {pages} {074023} (\bibinfo {year} {2016}{\natexlab{b}})},\ \Eprint {http://arxiv.org/abs/1607.00815} {arXiv:1607.00815 [hep-ph]} \BibitemShut {NoStop}%
\bibitem [{\citenamefont {Calibbi}\ \emph {et~al.}(2022)\citenamefont {Calibbi}, \citenamefont {Li}, \citenamefont {Marcano},\ and\ \citenamefont {Schmidt}}]{ref::psip2emu_significance}%
  \BibitemOpen
  \bibfield  {author} {\bibinfo {author} {\bibfnamefont {L.}~\bibnamefont {Calibbi}}, \bibinfo {author} {\bibfnamefont {T.}~\bibnamefont {Li}}, \bibinfo {author} {\bibfnamefont {X.}~\bibnamefont {Marcano}}, \ and\ \bibinfo {author} {\bibfnamefont {M.~A.}\ \bibnamefont {Schmidt}},\ }\href {\doibase 10.1103/PhysRevD.106.115039} {\bibfield  {journal} {\bibinfo  {journal} {Phys. Rev. D}\ }\textbf {\bibinfo {volume} {106}},\ \bibinfo {pages} {115039} (\bibinfo {year} {2022})},\ \Eprint {http://arxiv.org/abs/2207.10913} {arXiv:2207.10913 [hep-ph]} \BibitemShut {NoStop}%
\end{thebibliography}%

\vspace{0.8cm}
\noindent
\textbf{The BESIII experiment}
\vspace{0.2cm}

BESIII is an experiment located at the Beijing Electron Positron Collider (BEPCII) for the studies of hadron physics and $\tau$-charm physics. The BESIII detector consists of a helium-based multi-layer drift chamber, a plastic scintillator time-of-flight system, a CsI(Tl) electromagnetic calorimeter, and a muon identification system. The BESIII Collaboration has more than 500 members from 91 institutions in 15 countries.

\author{Author list}
\begin{small}
\vspace{0.4cm}
\noindent
\textbf{BESIII Collaboration}
\vspace{0.1cm}
\begin{center}
M.~Ablikim$^{1}$\BESIIIorcid{0000-0002-3935-619X},
M.~N.~Achasov$^{4,b}$\BESIIIorcid{0000-0002-9400-8622},
P.~Adlarson$^{77}$\BESIIIorcid{0000-0001-6280-3851},
X.~C.~Ai$^{82}$\BESIIIorcid{0000-0003-3856-2415},
R.~Aliberti$^{36}$\BESIIIorcid{0000-0003-3500-4012},
A.~Amoroso$^{76A,76C}$\BESIIIorcid{0000-0002-3095-8610},
Q.~An$^{59,73,\dagger}$,
Y.~Bai$^{58}$\BESIIIorcid{0000-0001-6593-5665},
O.~Bakina$^{37}$\BESIIIorcid{0009-0005-0719-7461},
Y.~Ban$^{47,g}$\BESIIIorcid{0000-0002-1912-0374}
% ...
\\
\vspace{0.2cm}
(BESIII Collaboration)\\
\vspace{0.2cm} {\it
$^{1}$ Institute of High Energy Physics, Beijing 100049, People's Republic of China\\
% ...
$^{4}$ Budker Institute of Nuclear Physics SB RAS (BINP), Novosibirsk 630090, Russia\\
% ...
$^{36}$ Johannes Gutenberg University of Mainz, Johann-Joachim-Becher-Weg 45, D-55099 Mainz, Germany\\
$^{37}$ Joint Institute for Nuclear Research, 141980 Dubna, Moscow region, Russia\\
% ...
$^{47}$ Peking University, Beijing 100871, People's Republic of China\\
% ...
$^{58}$ Southeast University, Nanjing 211100, People's Republic of China\\
$^{59}$ State Key Laboratory of Particle Detection and Electronics, Beijing 100049, Hefei 230026, People's Republic of China\\
% ...
$^{73}$ University of Science and Technology of China, Hefei 230026, People's Republic of China\\
% ...
$^{76}$ University of Turin and INFN, (A)University of Turin, I-10125, Turin, Italy; (B)University of Eastern Piedmont, I-15121, Alessandria, Italy; (C)INFN, I-10125, Turin, Italy\\
$^{77}$ Uppsala University, Box 516, SE-75120 Uppsala, Sweden\\
% ...
$^{82}$ Zhengzhou University, Zhengzhou 450001, People's Republic of China\\

\vspace{0.2cm}
$^{\dagger}$ Deceased\\
% ...
$^{b}$ Also at the Novosibirsk State University, Novosibirsk, 630090, Russia\\
% ...
$^{g}$ Also at State Key Laboratory of Nuclear Physics and Technology, Peking University, Beijing 100871, People's Republic of China\\
% ...

}
\end{center}

\vspace{0.4cm}
\end{small}

\end{document}